\documentclass[11pt]{article}

\usepackage{mathtools}
\usepackage[margin=1in]{geometry}
\usepackage{amsmath,amssymb,amsfonts}
\usepackage{enumitem}
\usepackage{amsthm}
\usepackage{xurl}
\usepackage[backend=biber,style=alphabetic,sorting=anyt,maxcitenames=3,mincitenames=1,maxalphanames=4,minalphanames=3,maxbibnames=99]{biblatex}
\usepackage{hyperref}
\hypersetup{
    colorlinks=true,
    linkcolor=blue,
    citecolor=blue,
    urlcolor=blue
}
\usepackage[nameinlink,capitalize]{cleveref}
\usepackage{thmtools}

\newtheorem{theorem}{Theorem}[section]

\newtheorem{remark}[theorem]{Remark}
\newtheorem{lemma}[theorem]{Lemma}
\newtheorem{proposition}[theorem]{Proposition}
\newtheorem{corollary}[theorem]{Corollary}
\newtheorem{definition}[theorem]{Definition}

\newcommand{\Dcal}{\mathcal{D}}
\newcommand{\Pcal}{\mathcal{P}}
\newcommand{\Vcal}{\mathcal{V}}

\newcommand{\dist}{\operatorname{dist}}
\newcommand{\TV}{\operatorname{TV}}
\newcommand{\supp}{\operatorname{supp}}
\newcommand{\viol}{\operatorname{viol}}
\newcommand{\cdist}{\operatorname{cdist}}
\newcommand{\packmod}{\Delta}
\newcommand{\compmod}{\Gamma}

\begin{document}

\hypersetup{pageanchor=false}
\begin{titlepage}
\title{Metric Testing: A Characterization of One-Sided Testability}
\author{Zihan Tan\thanks{University of Minnesota Twin Cities, MN, USA. Email: {\tt ztan@umn.edu}}
\and 
Theo Teske\thanks{University of Minnesota Twin Cities, MN, USA. Email: {\tt teske096@umn.edu}}
}

\maketitle
\thispagestyle{empty}
\begin{abstract}

We give a unified framework for one-sided testability (constant-query testing) of metric families. For a broad class of families defined by fixed local predicates, we prove that testability is equivalent to two structural properties: \emph{packing amplification}, which turns many edge-disjoint local violations into a positive density of violations, and \emph{completion stability}, which allows locally consistent partial matrices with few missing entries to be cheaply completed. This characterization yields a common proof strategy and simpler proofs of classical testability results for ultrametrics, tree pseudometrics, Euclidean pseudometrics in fixed dimension, and several others.

Our framework also gives new quantitative bounds for metric testing: a nonadaptive one-sided upper bound for matrices with entries in the bounded range $[1,C]$, and two-sided nonadaptive and adaptive lower bounds for general metrics.
\end{abstract}
\end{titlepage}
\hypersetup{pageanchor=true}

\tableofcontents

\newpage

\section{Introduction}

Given query access to a symmetric matrix $D\in \mathbb{R}^{n\times n}$ of nonnegative entries with zero diagonal and a prescribed metric family $\Pcal=\{\Pcal_n\}$, metric property testing asks us to distinguish whether $D\in\Pcal_n$ or $D$ is $\epsilon$-far from $\Pcal_n$.
Here, $\epsilon$-far means that more than $\epsilon n^2/2$ unordered
off-diagonal pairs must be changed to obtain a member of $\Pcal_n$.
The goal is to test the $\Pcal$-property using as few queries to the
entries of $D$ as possible.
Classic examples include testing whether $D$ satisfies the triangle inequality,
is an ultrametric, or can be realized by shortest-path distances in a tree or a
Euclidean space of fixed dimension~\cite{PR,KS,Onak,BKW}.
These questions have found numerous applications in
metric recognition, completion, and
repair~\cite{Menger,Schoenberg,Buneman,BJ,BDST,GJ,FRV,FGRSV,CFLM,Aranda}.

A \emph{one-sided $\epsilon$-tester} is a randomized algorithm that
accepts every member in $\Pcal_n$ with probability $1$ and rejects every
$\epsilon$-far input with probability at least $2/3$.
We call the family $\Pcal$ \emph{one-sided testable} if, for every $\epsilon>0$,
there is such a tester whose query complexity is bounded by
$q(\epsilon)$ independently of $n$.

Parnas and Ron~\cite{PR} initiated the study of metric property testing
and proved one-sided testability for several families,
including ultrametrics and non-injective tree and Euclidean metrics.
Krauthgamer and Sasson~\cite{KS} improved the testing bounds for finite
dimensional Euclidean realizability.
More recently, Bao, Kannan, and Waingarten~\cite{BKW} obtained improved
bounds for ultrametric testing, studied tree metric testing, and
gave a sublinear-query tester for general metrics.
They also proved lower bounds for general metric testing, ruling out nonadaptive one-sided
testers whose query complexity is polynomial in $1/\epsilon$
and independent of $n$.
Together, these results raise a structural question:

\begin{center}
\emph{What structural properties make a metric family one-sided testable?}
\end{center}

In fact, one-sided testability already forces locality: $\epsilon$-far
inputs must reveal small certificates of nonmembership under random sampling.
Indeed, for families invariant under relabelling, an adaptation of
the canonicalization argument of Goldreich and Trevisan~\cite{GT}
simulates a $q(\epsilon)$-query tester using the induced matrix on
a uniformly sampled set of at most $2q(\epsilon)$ points.
Perfect completeness forces the simulation to accept every sampled
configuration that occurs in a valid input.
Thus, on an $\epsilon$-far input, the sample must contain, with
probability at least $2/3$, a configuration incompatible with every
valid completion on $n$ points.
Small, detectable obstructions are therefore a necessary consequence
of one-sided testability.
This perspective also underlies Alon and Shapira's characterization
of oblivious one-sided graph testing through semi-hereditary
properties~\cite{AS}.

It is therefore natural to focus on metric families characterizable
by a \emph{local predicate}: a matrix belongs to the family
exactly when every $k$-point restriction satisfies the predicate,
where $k$ depends only on the family.
This structure is shared by many families:
general metrics and ultrametrics are determined by $3$-point conditions,
tree realizability by conditions on at most $4$ points, and
Euclidean realizability in fixed dimension by conditions on a bounded
number of points.
For these families, the challenge is to understand when being far
from membership makes local violations detectable by sampling.

Our main contribution is a unified structural characterization of
one-sided testability.
Our approach connects two tasks, repairing an input and detecting
its violations, through two complementary properties:
\begin{itemize}
    \item \emph{Packing amplification}: many independent violations force abundant violations.
    \item \emph{Completion stability}: few holes in locally consistent data require few edits.
\end{itemize}
To see how these properties work together, delete the pairs covered
by a maximal collection of edge-disjoint local violations.
Maximality ensures that the remaining partial matrix is locally
consistent.
If the collection is small, only a few holes are created, and
completion stability yields a valid matrix close to the original input.
Thus, an input far from the family must contain many edge-disjoint
violations, which packing amplification turns into sufficiently many violations
for random sampling to detect.
We show that these two properties together characterize one-sided
testability.
Separating these tasks gives a common proof strategy across metric
families, yields simplified proofs of several classical testability
results, and identifies distinct mechanisms behind the failure
of testability.

\subsection{Our results}

Fix a local predicate $\Phi$ on $k$ points characterizing a metric family $\Pcal$.
A $k$-point restriction that fails $\Phi$ is called a \emph{local violation}.
We introduce two properties, with all bounds uniform in $n\ge k$.

\emph{Packing amplification.}
For every $\alpha>0$, there exists $\delta>0$ such that any
$n$-point matrix containing at least $\alpha n^2$ edge-disjoint
local violations contains at least $\delta\binom{n}{k}$ local
violations in total.
Here, edge-disjointness means that no two violations share an
unordered pair of indices.

\emph{Completion stability.}
A partial matrix contains unspecified pairs, called \emph{holes}, and is
\emph{locally consistent} if every fully specified $k$-point
restriction satisfies $\Phi$.
For every $\eta>0$, there exists $\theta>0$ such that any locally
consistent partial matrix containing at most $\theta n^2$ holes can be
completed to a member of $\Pcal_n$ after changing at most
$\eta n^2$ specified pairs (filling the holes incurs no cost).

\begin{theorem}[Informal]
\label{thm:intro-characterization}
For metric families characterized by a local predicate, one-sided
testability is equivalent to packing amplification together with
completion stability.
\end{theorem}

Formal definitions and the precise theorem statement are given in \Cref{sec:structure}.

\begin{table}[htbp]
\centering
\small
\setlength{\tabcolsep}{4pt}
\renewcommand{\arraystretch}{1.3}
\begin{tabular}{|l|c|c|c|}
\hline
Family & Packing amplification & Completion stability & Previous work \\
\hline
Ultrametrics
& Yes (Lemma~\ref{lem:ult-packing})
& Yes (Lemma~\ref{lem:ult-completion})
& \cite{PR} \\
Strict ultrametrics
& Yes (Corollary~\ref{cor:strict-packing})
& Yes (Corollary~\ref{cor:strict-ult-completion})
& \cite{BKW} \\
Tree pseudometrics
& Yes (Lemma~\ref{lem:tree-packing})
& Yes (Lemma~\ref{lem:tree-completion-stability})
& \cite{PR} \\
Euclidean pseudometrics in $\mathbb{R}^d$
& Yes (Lemma~\ref{lem:packing-euclidean-pseudo})
& Yes (Lemma~\ref{lem:completion-stability-euclidean-pseudo})
& \cite{PR,KS} \\
\hline
Strict tree metrics
& Yes (Corollary~\ref{cor:strict-packing})
& No (\Cref{thm:completion-obstructions})
& \cite{PR} \\
Strict Euclidean metrics in $\mathbb{R}^d$
& Yes (Corollary~\ref{cor:strict-packing})
& No (\Cref{thm:completion-obstructions})
& \cite{PR} \\
Bounded aspect ratio metrics
& No (\Cref{thm:metric-packing-obstruction})
& Yes (\Cref{thm:completion-stability-C})
& --- \\
General metrics
& No (\Cref{thm:metric-packing-obstruction})
& ---
& \cite{BKW} \\
\hline
\end{tabular}
\caption{Packing amplification and completion stability for the metric families studied in this paper. The dimension $d\geq 1$ is a fixed constant.
Strictness requires positive distances between distinct labels. For metrics with bounded aspect ratio, off-diagonal entries lie in $[1,C]$ for a fixed $C>2$.}
\label{tab:intro-classification}
\end{table}

\paragraph{Unified and simplified analysis for one-sided testability.}
We then apply this characterization to several metric families (see Table~\ref{tab:intro-classification}).
Our framework recovers the qualitative one-sided testability results of~\cite{PR,KS,BKW} for ultrametrics, strict ultrametrics, tree pseudometrics, and Euclidean pseudometrics in fixed dimension, with unified and simpler proofs via the packing amplification and the completion stability properties. For ultrametrics, these properties follow from elementary counting and partitioning arguments. For tree pseudometrics, both properties reduce to their ultrametric counterparts.
For Euclidean pseudometrics of fixed dimension, both arguments use the rigidity of distances to a small affine anchor set.
Therefore, the same testing argument applies across families, and parts of the proofs can be reused.
The same framework identifies distinct reasons for non-testability.
General metrics and metrics with bounded aspect ratio (entries in
$[1,C]$ for fixed $C>2$) fail packing amplification.
Strict tree and Euclidean metrics retain packing amplification
but fail completion stability.
Also, as metrics with bounded aspect ratio satisfy completion stability, these examples establish the independence of the two properties.

\paragraph{Quantitative testing bounds.}
Beyond giving a qualitative classification, our structural framework also gives new quantitative upper and lower bounds. Because our characterization reduces one-sided testability to two combinatorial properties, bounding completion stability and packing amplification yields explicit query upper bounds. For ultrametrics, strict ultrametrics, and tree pseudometrics, combining linear completion stability with quadratic packing amplification yields one-sided $O(\epsilon^{-2})$-query testers (Corollary~\ref{cor:geometric-testing}), removing the logarithmic factors from the $\widetilde{O}(\epsilon^{-2})$ bounds of~\cite{BKW}. We also provide novel bounds for general $n$-point metrics and metrics with entries in $[1,C]$. To state the full tradeoff across all proximity regimes, define the rate function
$$
F(n,\epsilon)=
\begin{cases}
n^2, & 0<\epsilon\le n^{-2},\\[1mm]
n/\sqrt{\epsilon}, & n^{-2}<\epsilon\le n^{-1},\\[1mm]
\sqrt n/\epsilon, & n^{-1}<\epsilon\le n^{-1/2},\\[1mm]
(n/\epsilon)^{2/3}, & n^{-1/2}<\epsilon\le 1.
\end{cases}
$$

\begin{theorem}
\label{thm:intro-metric-bounds}
Fix $C>2$ and $\epsilon_0=1/80$. For all sufficiently large $n$ and $0<\epsilon\le\epsilon_0$:
\begin{enumerate}[label=(\arabic*)]
    \item \emph{Sharp nonadaptive complexity on $[1,C]$ (\Cref{thm:sharp-nonadaptive-metric}):} For inputs with entries in $[1,C]$, both one-sided and two-sided nonadaptive $\epsilon$-testing of general metrics have query complexity $\Theta_C(F(n,\epsilon))$.
    \item \emph{Adaptive lower bound on $[1,C]$ (\Cref{thm:metric-adaptive}):} Any adaptive two-sided $\epsilon$-tester for general metrics on $[1,C]$ requires $\Omega_C(\sqrt{n/\log n})$ queries.
    \item \emph{Transfer to general and $\delta$-hyperbolic metrics:} On arbitrary nonnegative inputs, nonadaptive and adaptive two-sided $\epsilon$-testing require $\Omega(F(n,\epsilon))$ and $\Omega(\sqrt{n/\log n})$ queries, respectively, both for general metrics and pseudometrics (\Cref{thm:unrestricted-metric-lower}) and for $\delta$-hyperbolic metrics with any fixed $\delta>0$ (Remark~\ref{rem:hyperbolic-packing}).
\end{enumerate}
\end{theorem}

Previously, Bao, Kannan, and Waingarten~\cite{BKW} gave a nonadaptive one-sided query lower bound of $\Omega(n^{2/3-o(1)})$ using Behrend graphs that applied only at vanishing proximity $\epsilon = n^{-o(1)}$. Because our packing obstruction avoids progression-free sets, \Cref{thm:intro-metric-bounds} eliminates the $n^{o(1)}$ loss and establishes hardness at fixed constant $\epsilon$ even against two-sided and adaptive testers.

\subsection{Related work, and comparison with our results}

Parnas and Ron~\cite{PR} first introduced metric property testing, giving one-sided testers with $O(\epsilon^{-6})$ queries for tree pseudometrics and ultrametrics, and
$O(d^2\log^2(d+1)/\epsilon^2)$ queries for Euclidean pseudometrics in $\mathbb{R}^d$.
Krauthgamer and Sasson~\cite{KS} improved the latter bound to
$O(d^2/\epsilon^2)$ using $O(d/\epsilon)$ sampled points.
For strict tree metrics and injective Euclidean realizations, Parnas and Ron proved $\Omega(\sqrt n)$ query lower bounds even with two-sided error. Our proofs unify these qualitative results; our Euclidean bound has weaker $\epsilon$-dependence than~\cite{KS}.

Bao, Kannan, and Waingarten~\cite{BKW} established a nonadaptive one-sided
$O(n^{2/3}/\epsilon^{4/3})$-query tester for general metrics and improved the ultrametric testing bound to $\widetilde O(\epsilon^{-2})$. They also studied tree metric testing.
Their Behrend graph lower bound for general metrics applies at $\epsilon$ vanishing with $n$.
Our obstruction holds at fixed $\epsilon$, even for entries
in $[1,C]$ with any fixed $C>2$.
Onak~\cite{Onak} studied tree metrics, ultrametrics, and line
embeddings in a model measuring distance by point modifications.

For graphs, Goldreich and Trevisan~\cite{GT} canonicalized
constant-query testers using uniformly sampled induced subgraphs.
Alon and Shapira~\cite{AS} established one-sided testability of every
hereditary graph property and characterized oblivious one-sided
testability.
Austin and Tao~\cite{AT} developed testing and repair results for
hereditary graph and hypergraph properties; Alon, Ben-Eliezer,
and Fischer~\cite{ABF} established testability for hereditary
ordered graphs and finite-alphabet matrices.
We decompose removability for real-valued matrices into two structural
conditions; completion stability concerns existence of close
completions and differs from the local repairability in~\cite{AT}.

Metric repair and the metric violation distance problem also aim at minimizing the number of changed entries~\cite{GJ,FRV} of an input matrix.
Fan, Gilbert, Raichel, Sonthalia, and Van Buskirk~\cite{FGRSV} studied incomplete graphs using
nonmetric cycles.
Cohen-Addad, Fan, Lee, and de Mesmay~\cite{CFLM} obtained an $O(\log n)$-approximation for metric violation distance and a constant-factor approximation for ultrametric violation distance.
Bandyapadhyay, Lochet, Lokshtanov, Saurabh, Sonar, Surianarayanan, and Xue~\cite{BLLSSSX} gave a single-exponential
fixed-parameter algorithm and a linear kernel for metric repair,
and proved fixed-parameter tractability of tree metric repair.
Our bounded range completion theorem controls the edits needed
to fill sparse holes, including those forced by longer nonmetric
cycles.
Metric nearness instead minimizes a norm of the numerical
perturbation~\cite{BDST}.

\subsection{Organization of the paper}

\Cref{sec:prelim} introduces the testing model, local predicates, and the metric families considered in the paper. \Cref{sec:structure} proves the structural characterization of one-sided testability through removal, packing amplification, and completion stability.
\Cref{sec:packing} establishes packing amplification for the geometric families and constructs a packing obstruction for general metrics. \Cref{sec:completion} proves the corresponding completion results and identifies the obstruction caused by injective realizations. Finally, \Cref{sec:quantitative-testing} derives quantitative testing bounds, including sharp nonadaptive bounds for metrics with bounded aspect ratio and an adaptive lower bound.

\section{Preliminaries}\label{sec:prelim}

Fix an integer $k\ge2$ and a nonempty value set $\Lambda\subseteq\mathbb R_{\ge0}$. For $n\ge1$, let $\Dcal_n$ be the set of symmetric matrices with off-diagonal entries in $\Lambda$ and diagonal entries fixed to $0$.

For $A,B\in\Dcal_n$, define the \emph{normalized Hamming distance} between $A,B$ as
$$
\dist_n(A,B)
=
\frac{2}{n^2}
\left|\left\{e\in\binom{[n]}2:A_e\ne B_e\right\}\right|.
$$
For a nonempty set $\mathcal Q\subseteq\Dcal_n$, define $\dist_n(A,\mathcal Q)=\min_{B\in\mathcal Q}\dist_n(A,B)$.

For probability distributions $\mu,\nu$, denote by $\TV(\mu,\nu)=\sup_E|\mu(E)-\nu(E)|$ the total variation distance between $\mu,\nu$, where the supremum is over measurable events.


\subsection*{Local Predicates and Testability}

Let $\Phi\subseteq\Lambda^{\binom{[k]}2}$ be a permutation-invariant $k$-point predicate. 
For $S\in\binom{[n]}k$, membership of $D[S]$ in $\Phi$ is interpreted after pulling $D[S]$ back along any bijection $[k]\to S$.

Define
$$
\Pcal_n
=
\left\{D\in\Dcal_n:
D[S]\text{ satisfies }\Phi
\text{ for every }S\in\binom{[n]}k
\right\}.
$$
For $n<k$, $\Pcal_n=\Dcal_n$. Unless stated otherwise, we assume $\Pcal_n\ne\varnothing$ for every $n\ge k$.


For $D\in\Dcal_n$ with $n\ge k$, define
$$
\Vcal(D)
=
\left\{S\in\binom{[n]}k:D[S]\text{ violates }\Phi\right\},
\qquad
\text{and}
\qquad
\viol(D)=\frac{|\Vcal(D)|}{\binom nk}.
$$


\begin{definition}[One-sided tester]
A one-sided $q$-query $\epsilon$-tester for $\Pcal=\{\Pcal_n\}$ is an information-theoretic randomized adaptive decision tree which queries at most $q$ off-diagonal entries and satisfies:
\begin{enumerate}[label=(\arabic*)]
\item every $D\in\Pcal_n$ is accepted with probability $1$;
\item every $D$ with $\dist_n(D,\Pcal_n)>\epsilon$ is rejected with probability at least $2/3$.
\end{enumerate}
\end{definition}

If the query bound $q=q(\epsilon)$ in the above definition depends only on $\Pcal$ and $\epsilon$ but not $n$, then we say that $\Pcal$ is \emph{one-sided testable}.

\begin{definition}[Two-sided tester]
A two-sided $q$-query $\epsilon$-tester for $\Pcal=\{\Pcal_n\}$ is an information-theoretic randomized adaptive decision tree which queries at most $q$ off-diagonal entries and satisfies:
\begin{enumerate}[label=(\arabic*)]
\item every $D\in\Pcal_n$ is accepted with probability at least $2/3$;
\item every $D$ with $\dist_n(D,\Pcal_n)>\epsilon$ is rejected with probability at least $2/3$.
\end{enumerate}
\end{definition}




\subsection*{Metric families and their extendability}

\begin{definition}[Extendability]
A family $\Pcal=\{\Pcal_n\}$ is \emph{extendable} if, for every $n\ge r\ge k$ and every $A\in\Pcal_r$, there exist $B\in\Pcal_n$ and an injection $\iota:[r]\hookrightarrow[n]$ such that
$
B_{\iota(a)\iota(b)}=A_{ab}$
for every $ab\in\binom{[r]}2$.
\end{definition}

We now define the principal metric and geometric families and verify their extendability. To distinguish arbitrary pseudometric realizations from injective realizations, we use a superscript $+$ for predicates and families restricted to positive off-diagonal entries $\Gamma=(0,\infty)\subseteq\mathbb R_{\ge0}$.

The tree and Euclidean symbols below denote local families with the preceding vacuous convention at orders below the predicate arity. Identifications with global geometric realization families are asserted at and above that arity. If global realizability is required at the finitely many smaller orders, those inputs can instead be inspected completely; none of the asymptotic conclusions is affected.

\subsubsection*{General and bounded aspect ratio metrics}

Write $\mathcal M_n$ for the positive-valued metrics on $[n]$ and $\overline{\mathcal M}_n$ for the nonnegative-valued pseudometrics on $[n]$. For $C>1$, define
$
\mathcal M_n[1,C]
=
\mathcal M_n\cap[1,C]^{\binom{[n]}2}$.

On the ambient value set $\Lambda=[1,C]$ with $k=3$, the family $\mathcal M[1,C]=\{\mathcal M_n[1,C]\}$ is defined by
$$
\Phi_{\mathrm{met}}
=
\left\{
(a,b,c)\in[1,C]^3:
a\le b+c,\ b\le a+c,\ c\le a+b
\right\}.
$$
Metrics with bounded aspect ratio $\mathcal M[1,C]$ are extendable by assigning distance $C$ to every pair involving a new vertex. General metrics $\mathcal M$ and pseudometrics $\overline{\mathcal M}$ are extendable by assigning the common value
$
R=\max\bigl(\{1\}\cup\{\text{old entries}\}\bigr)$
to every pair involving a new vertex.

\subsubsection*{Ultrametrics}

Let $k=3$ and define
$$
\Phi_{\mathrm{ult}}
=
\{(a,b,c)\in\Lambda^3:
\max\{a,b,c\}\text{ is attained at least twice}\}.
$$
Denote by $\mathcal U=\{\mathcal U_n\}$ the family. For $\Lambda=\mathbb R_{\ge0}$, denote
$
\Phi_{\mathrm{ult}}^+
=
\Phi_{\mathrm{ult}}\cap(0,\infty)^3
$
and $\mathcal U^+=\{\mathcal U_n^+\}$ for the strict ultrametrics.

\begin{lemma}\label{lem:ult-extendability}
The family $\mathcal U$ is extendable over any nonempty value set $\Lambda$, and $\mathcal U^+$ is extendable.
\end{lemma}

\begin{proof}
Let $n\ge r\ge3$ and $A\in\mathcal U_r$. Set
$R=\max_{e\in\binom{[r]}2}A_e\in\Lambda$.
For all distinct pairs of indices $i,j\le r$, define $B_{ij}=A_{ij}$, otherwise define
$B_{ij}=R$.
Every triangle containing a new vertex has at least two edges of value $R$, and its remaining edge has value at most $R$. Thus $B\in\mathcal U_n$. If $A$ is strict, then $R>0$ and $B$ is strict as well.
\end{proof}

\subsubsection*{Tree pseudometrics and strict tree metrics}

Set $k=4$ and $\Lambda=\mathbb R_{\ge0}$. Define $\Phi_{\mathrm{tree}}$ to consist of those symmetric, zero-diagonal matrices $D$ on $[4]$ for which there exist a weighted tree $T=(V,E)$, positive edge weights $w:E\to(0,\infty)$, and a not necessarily injective map $\phi:[4]\to V$ satisfying
$D_{ij}=\dist_T(\phi(i),\phi(j))$ for all $i,j\in[4]$.

Equivalently, by Buneman's characterization \cite{Buneman}, $D$ satisfies the \emph{four-point condition}
$$
D_{ij}+D_{kl}
\le
\max\{D_{ik}+D_{jl},D_{il}+D_{jk}\}
\qquad(i,j,k,l\in[4]),
$$
with repeated indices allowed, thereby including triangle inequalities.

Let $\mathcal T=\{\mathcal T_n\}$ be the resulting local family. Define
$
\Phi_{\mathrm{tree}}^+
=
\Phi_{\mathrm{tree}}\cap(0,\infty)^{\binom{[4]}2},
$
and let $\mathcal T^+=\{\mathcal T_n^+\}$ be its local family. The strict predicate characterizes injective realizations in positively weighted trees, allowing labeled internal vertices and additional unlabeled vertices. By the classical local-to-global theorem \cite{Buneman}, for $n\ge4$, $\mathcal T_n$ and $\mathcal T_n^+$ coincide with the matrices globally realized in weighted trees by arbitrary and injective maps, respectively. Edges of length zero in a representation with nonnegative edge weights may be contracted; positivity of all off-diagonal distances prevents distinct labels from being identified in the strict case.

\begin{lemma}\label{lem:tree-extendability}
The families $\mathcal T$ and $\mathcal T^+$ are extendable.
\end{lemma}

\begin{proof}
Let $n\ge r\ge4$ and $A\in\mathcal T_r$. Realize $A$ by a weighted tree. Attach each new label to the point representing label $1$ by a new edge of length $1$. Equivalently, define
$$
B_{ij}
=
\begin{cases}
A_{ij},&i,j\in[r],\\
A_{i1}+1,&i\in[r],\ j\notin[r],\\
2,&i,j\notin[r],\ i\ne j,
\end{cases}
$$
and impose symmetry. This is a tree pseudometric preserving all old entries. If the old realization is injective, the new realization is injective as well.
\end{proof}

\subsubsection*{$d$-dimensional Euclidean metrics}

Fix an integer $d\ge1$. Denote $k=d+3$ and $\Lambda=\mathbb R_{\ge0}$. Define $\Phi_{\mathrm{Euc}}^{(d)}$ as the set of symmetric, zero-diagonal matrices $D$ on $[k]$ for which there exists a (not necessarily injective) mapping $\phi:[k]\to\mathbb R^d$ satisfying
$D_{ij}=\|\phi(i)-\phi(j)\|_2$ for all $i,j\in[k]$.
Let $\mathcal E^{(d)}=\{\mathcal E_n^{(d)}\}$ be the resulting local family. Define
$
\Phi_{\mathrm{Euc}}^{(d),+}
=
\Phi_{\mathrm{Euc}}^{(d)}
\cap(0,\infty)^{\binom{[k]}2},
$
and let $\mathcal E^{(d),+}=\{\mathcal E_n^{(d),+}\}$ be its local family.

The affine dimension of a Euclidean distance matrix is well-defined. Indeed, after choosing a label $o$, the Gram matrix $G$ where
$
G_{ij}
=
\frac{D_{oi}^2+D_{oj}^2-D_{ij}^2}{2}$ for all $i,j\ne o$
is determined by the distances, and its rank is the affine dimension of every realization. By the classical theorem of Schoenberg~\cite{Schoenberg} and Young and Householder~\cite{YH}, positive semidefiniteness and rank at most $d$ of $G$ characterize realization in $\mathbb R^d$. This also gives a rational decision procedure and Borel defining conditions.

Recall that vectors $z_0,\ldots,z_s$ are \emph{affinely independent} if
$\sum_{i=0}^s\alpha_i z_i=0$ and $\sum_{i=0}^s\alpha_i=0$
imply $\alpha_0=\cdots=\alpha_s=0$.

\begin{lemma}\label{lem:affine-basis-unique}
Let vectors $z_0,\ldots,z_d\in\mathbb R^d$ be affinely independent. If points $x,y\in\mathbb R^d$ satisfy
$\|x-z_i\|_2=\|y-z_i\|_2$
for all $i=0,\ldots,d$,
then $x=y$.
The same assertion holds inside the affine span of any affinely independent set, including a singleton.
\end{lemma}

\begin{proof}
Subtracting the squared-distance equality for $z_0$ from that for $z_i$ gives
$$
\langle x-y,z_i-z_0\rangle=0
\qquad(i=1,\ldots,d).
$$
These difference vectors span $\mathbb R^d$, so $x-y=0$. The affine-span version is the same argument after identifying the affine span with a Euclidean space; the singleton case is immediate.
\end{proof}

We record the finite local characterization of Menger \cite{Menger,Blumenthal}.

\begin{lemma}[Local characterization of Euclidean metrics]\label{lem:euclidean-local-char}
Let $n\ge d+3$ and $D\in\Dcal_n$. Suppose every $(d+3)$-point principal submatrix of $D$ is realized in $\mathbb R^d$. Then $D$ is realized in $\mathbb R^d$. If all off-diagonal entries of $D$ are positive, the realizing points are pairwise distinct.
\end{lemma}

\begin{proof}
Every principal submatrix of order at most $d+3$ is realizable, by extending its label set to a $(d+3)$-set.
Choose a set $Z\subseteq[n]$ of maximum cardinality such that $D[Z]$ has an affinely independent realization in $\mathbb R^d$. Write $|Z|=s+1\le d+1$, fix a realization
$
V=\{v_0,\ldots,v_s\},
$
and let $A$ be its affine span.

For $x\notin Z$, the matrix $D[Z\cup\{x\}]$ is realizable. Two realizations of the affine basis have the same Gram matrix and can be aligned by a Euclidean isometry. After aligning $Z$ with $V$, the realization of $x$ lies in $A$: otherwise $Z\cup\{x\}$ would be affinely independent, contrary to maximality. Its position in $A$ is unique by Lemma~\ref{lem:affine-basis-unique}. Denote it by $\phi(x)$, and use the prescribed points of $V$ for labels in $Z$.

For distinct $x,y\notin Z$, the set $Z\cup\{x,y\}$ has size at most $d+3$ and is realizable. Align its realization on $Z$. Maximality again places both remaining points in $A$, and uniqueness identifies them with $\phi(x)$ and $\phi(y)$. Hence
$
D_{xy}=\|\phi(x)-\phi(y)\|_2.
$
All other distances are preserved by construction. Positive off-diagonal distances imply injectivity.
\end{proof}

\begin{lemma}\label{lem:euclidean-extendability}
The families $\mathcal E^{(d)}$ and $\mathcal E^{(d),+}$ are extendable.
\end{lemma}

\begin{proof}
Let $n\ge r\ge d+3$ and $A\in\mathcal E_r^{(d)}$. By Lemma~\ref{lem:euclidean-local-char}, realize $A$ by points $x_1,\ldots,x_r\in\mathbb R^d$. Choose additional points that are pairwise distinct and distinct from every old point. Their Euclidean distance matrix extends $A$ and belongs to $\mathcal E_n^{(d)}$. If $A$ was strict, all realizing points are distinct, so the extension is strict.
\end{proof}

\section{Characterization of one-sided testing: Proof of Theorem~\ref{thm:intro-characterization}}\label{sec:structure}

We characterize constant-query one-sided testability for extendable local matrix properties. We begin by showing that one-sided testability is equivalent to removability via a canonicalization framework (Theorem~\ref{thm:testing}). We then establish our main structural characterization (Theorem~\ref{thm:structure}), proving that removability decomposes into two combinatorial conditions: packing amplification and completion stability. Together, these results yield our main equivalence (Corollary~\ref{cor:test-structure}) and isolate the two structural obstructions to testability (Corollary~\ref{cor:obstructions}).

\subsection{Canonical testers and removability}

To prove Theorem~\ref{thm:testing}, we first reduce arbitrary adaptive one-sided testers to canonical testers that inspect a uniformly random induced submatrix. The lazy permutation coupling in Lemma~\ref{lem:lazy} and the resulting canonical reduction in Lemma~\ref{lem:canonical-domination} and Theorem~\ref{thm:testing} adapt the canonicalization framework of Goldreich and Trevisan~\cite[Section~4; cf.\ Lemma~4.1 and Theorem~4.5]{GT} from graph properties to extendable local matrix properties. This connects one-sided testability to removal lemmas for hereditary properties in the spirit of Alon and Shapira~\cite{AS}.

\begin{lemma}\label{lem:lazy}
Let $n\ge r$. Fix an adaptive decision tree whose queries are unordered pairs of labels from $[n]$ and which touches at most $r$ distinct labels. For a fixed input $D\in\Dcal_n$, compare:
\begin{enumerate}[label=(\arabic*)]
\item choose a uniform permutation $\sigma$ of $[n]$, and answer $\{i,j\}$ with $D_{\sigma(i)\sigma(j)}$;
\item choose a uniform $r$-set $U\subseteq[n]$, assign each newly touched label to a uniformly random unused element of $U$, and answer using the resulting partial injection.
\end{enumerate}
The answer transcripts have the same distribution. Moreover, the experiments can be coupled to have the same sequence of images of newly touched labels.
\end{lemma}

\begin{proof}
Expose newly touched labels in a fixed order when a query introduces two of them. Suppose the exposed images are distinct vertices $v_1,\ldots,v_s$, where $s<r$.

In the permutation experiment, the next image is uniform on the unused vertices of $[n]$. In the $r$-set experiment, conditional on these assignments, $U$ is uniform among the $r$-sets containing $\{v_1,\ldots,v_s\}$: for each such $U$, the probability of choosing $U$ and making these assignments is $\binom nr^{-1}\frac{(r-s)!}{r!}$. Thus the next image is again uniform on the unused vertices of $[n]$.

For a fixed input and decision tree, previous answers and subsequent query choices are determined by the exposed images. Conditioning on the complete partial history therefore adds no restriction on the remaining assignments. Induction proves the claim.
\end{proof}

\begin{lemma}\label{lem:canonical-domination}
Assume $\Pcal$ is extendable. Let $A$ be a one-sided $q$-query $\epsilon$-tester, and put $r=\max\{k,2q\}$. For $n\ge r$, symmetrize $A$ by a uniform permutation of $[n]$. Let $C$ query a uniform $r$-set $U$ completely and lazily simulate $A$ within $U$. Then
$$
\Pr[C\text{ rejects }D]
=
\Pr[A^{\mathrm{sym}}\text{ rejects }D]
\le
\Pr_U[D[U]\notin\Pcal_r].
$$
\end{lemma}

\begin{proof}
The equality follows from Lemma~\ref{lem:lazy}, conditional on the internal randomness of $A$.

Fix $U$ with $D[U]\in\Pcal_r$. Extendability gives a valid $B\in\Pcal_n$ containing an induced copy of $D[U]$ on some $r$-set $W$. Conditional on sampling $U$ in $D$ or $W$ in $B$, the lazy simulations have the same transcript law. Since $W$ has positive sampling probability, positive conditional rejection probability on $W$ would contradict one-sided completeness on $B$. Thus the conditional rejection probability on $U$ is zero.
\end{proof}

\begin{definition}[Removal modulus]
For $\epsilon>0$, define
$$
\rho(\epsilon)=  \inf_{\substack{n\ge k,\ D\in\Dcal_n\\ \dist_n(D,\Pcal_n)>\epsilon}} \viol(D).
$$
We say that $\Phi$ is \emph{removable} if $\rho(\epsilon)>0$ for every $\epsilon>0$.

Equivalently, for every $\epsilon>0$, there exists $\delta(\epsilon)>0$ such that
$$
\viol(D)<\delta(\epsilon)
\quad\Longrightarrow\quad
\dist_n(D,\Pcal_n)\le\epsilon
$$
for all $n\ge k$ and $D\in\Dcal_n$.
\end{definition}

\begin{theorem}[Testing-removal equivalence]\label{thm:testing}
Assume $\Phi$ is measurable, and fix $\epsilon>0$. If $\rho(\epsilon)>0$, then $\Pcal$ has a one-sided $\epsilon$-tester using at most $\binom k2 \bigl\lceil\frac{\log3}{\rho(\epsilon)}\bigr\rceil$ queries.

Conversely, suppose $\Pcal$ is extendable and has a one-sided $q$-query $\epsilon$-tester. Put $r=\max\{k,2q\}$. Querying a uniform $r$-submatrix $D[U]$ and rejecting if $D[U]\notin\Pcal_r$ is a one-sided $\epsilon$-tester using at most $\binom r2$ queries; when $n<r$, query the whole matrix instead. Moreover,
$$
\rho(\epsilon)\ge\frac{2}{3\binom rk}.
$$
Consequently, under extendability, removability is equivalent to one-sided testability with query complexity independent of $n$.
\end{theorem}

\begin{proof}
For the forward implication, independently sample $s=\bigl\lceil\frac{\log3}{\rho(\epsilon)}\bigr\rceil$ uniform $k$-sets and reject if one violates $\Phi$. An $\epsilon$-far input is accepted with probability at most $(1-\rho(\epsilon))^s\le e^{-\rho(\epsilon)s}\le 1/3$.

For the converse, Lemma~\ref{lem:canonical-domination} implies that an $\epsilon$-far input of order $n\ge r$ has an invalid uniform $r$-submatrix with probability at least $2/3$. Hence
$$
\frac23
\le
\sum_{T\in\Vcal(D)}\Pr[T\subseteq U]
=
|\Vcal(D)|\frac{\binom rk}{\binom nk}
=
\binom rk\,\viol(D).
$$
For $k\le n<r$, an $\epsilon$-far input has at least one violation, so $\viol(D)\ge\binom nk^{-1}\ge\binom rk^{-1}$. Taking the infimum proves the bound. Measurability makes the local membership decisions admissible.
\end{proof}

\subsection{Structural decomposition of removability}

We now introduce the two combinatorial properties that govern removability: packing amplification and completion stability.

\begin{definition}[Packing amplification]
A family $\mathcal S\subseteq\Vcal(D)$ is \emph{edge-disjoint} if no edge belongs to more than one member. Let $\nu(D)$ be its maximum possible size. For $\alpha>0$, define
$$
\packmod(\alpha)
=
\inf_{\substack{n\ge k,\ D\in\Dcal_n\\
\nu(D)\ge\alpha n^2}}
\viol(D).
$$
We say that $\Phi$ has \emph{packing amplification} if $\packmod(\alpha)>0$ for every $\alpha>0$.
\end{definition}

\begin{definition}[Completion stability]
A partial matrix $A$ on $[n]$ assigns values in $\Lambda$ to some pairs. Its set of unspecified edges is denoted $H(A)$; these edges are called \emph{holes}. The partial matrix is \emph{locally consistent} if every fully specified $k$-subset satisfies $\Phi$. For such a partial matrix with $n\ge k$, define
$$
\cdist(A)
=
\frac{2}{n^2}
\min_{B\in\Pcal_n}
\left|
\left\{
e\in\binom{[n]}2\setminus H(A):B_e\ne A_e
\right\}
\right|.
$$
For $\theta\ge0$, define $\compmod(\theta)=\sup_A\cdist(A)$, where the supremum is over locally consistent partial matrices on $n\ge k$ vertices with $|H(A)|\le\theta n^2$. The function $\compmod$ is non-decreasing. We say that $\Phi$ is \emph{completion stable} if $\compmod(\theta)\to0$ as $\theta\downarrow0$.
\end{definition}

\begin{lemma}\label{lem:packing-viol-edits}
For every $D\in\Dcal_n$ with $n\ge k$, we have $\dist_n(D,\Pcal_n)\ge 2\nu(D)/n^2$. Consequently, $\packmod(\alpha)\ge\rho(\alpha)$ for all $\alpha>0$.
\end{lemma}

\begin{proof}
A valid repair changes at least one edge in each member of an edge-disjoint family of violations, and these edits are distinct. If $\nu(D)\ge\alpha n^2$, then $\dist_n(D,\Pcal_n)\ge2\alpha>\alpha$, so $\viol(D)\ge\rho(\alpha)$. Take the infimum.
\end{proof}

\begin{lemma}\label{lem:holes-limit-viol}
Let $A$ be locally consistent on $n\ge k$ vertices, and let $D$ be any completion of $A$. If $|H(A)|\le\theta n^2$, then $\viol(D)\le4\binom k2\theta$.
\end{lemma}

\begin{proof}
Every violating $k$-set contains a hole, so $|\Vcal(D)|\le |H(A)|\binom{n-2}{k-2}$. Thus $\viol(D) \le |H(A)|\frac{k(k-1)}{n(n-1)} \le 2\binom k2\theta\frac n{n-1} \le 4\binom k2\theta$.
\end{proof}

\begin{proposition}[Deterministic repair inequality]\label{prop:repair-ineq}
For every $D\in\Dcal_n$ with $n\ge k$,
$$
\dist_n(D,\Pcal_n)
\le
\compmod\!\left(\frac{\binom k2\nu(D)}{n^2}\right)
+
\frac{2\binom k2\nu(D)}{n^2}.
$$
\end{proposition}

\begin{proof}
Take a maximum edge-disjoint family $\mathcal S$ of violations and delete all its edges. The resulting hole set has size at most $\binom k2\nu(D)$. The remaining partial matrix $A$ is locally consistent, since a fully specified violation would be edge-disjoint from $\mathcal S$.

A valid matrix attaining $\cdist(A)$ differs from $D$ on specified edges by at most $\compmod(\binom k2\nu(D)/n^2)$ and on holes by at most $2\binom k2\nu(D)/n^2$. Adding the two bounds completes the proof.
\end{proof}

\begin{theorem}[Structural characterization of removability]\label{thm:structure}
The predicate $\Phi$ is removable if and only if it is completion stable and has packing amplification.
\end{theorem}

\begin{proof}
Suppose $\Phi$ is removable. Fill the holes of a locally consistent partial matrix $A$ arbitrarily. If its hole density is at most $\theta$, Lemma~\ref{lem:holes-limit-viol} implies that the resulting full matrix has violation density at most $4\binom k2\theta$. Whenever $4\binom k2\theta<\rho(\epsilon)$, it has a valid repair at distance at most $\epsilon$. Restricting discrepancies to the originally specified edges gives $\cdist(A)\le\epsilon$, so $\compmod(\theta)\to0$. Packing amplification follows from Lemma~\ref{lem:packing-viol-edits}.

Conversely, choose $\theta>0$ such that $\compmod(\theta)+2\theta\le\epsilon$. If $D$ is $\epsilon$-far, Proposition~\ref{prop:repair-ineq} implies $\nu(D)>\theta n^2/\binom k2$. Therefore $\viol(D)\ge \packmod(\theta/\binom k2)$, and taking the infimum gives the same positive lower bound for $\rho(\epsilon)$. This remains valid if there are no $\epsilon$-far inputs.
\end{proof}

\begin{corollary}[Testing-structure equivalence]\label{cor:test-structure}
If $\Pcal$ is extendable and $\Phi$ is measurable, the following are equivalent:
\begin{enumerate}[label=(\arabic*)]
\item $\Pcal$ has a one-sided constant-query $\epsilon$-tester for every $\epsilon>0$;
\item $\Phi$ is removable;
\item $\Phi$ is completion stable and has packing amplification.
\end{enumerate}
\end{corollary}

\begin{proof}
Apply Theorems~\ref{thm:testing} and~\ref{thm:structure}.
\end{proof}

\begin{corollary}[Two obstructions to removal]\label{cor:obstructions}
The predicate $\Phi$ is not removable if and only if at least one of the following exists:
\begin{enumerate}[label=(\arabic*)]
\item \emph{Packing obstruction.} There are $\alpha>0$ and matrices $A^{(j)}$ on $s_j\to\infty$ vertices such that $\nu(A^{(j)})\ge\alpha s_j^2$ and $\viol(A^{(j)})\to0$.
\item \emph{Completion obstruction.} There are $\gamma>0$ and locally consistent partial matrices $A^{(j)}$ on $s_j\to\infty$ vertices such that $|H(A^{(j)})|/s_j^2\to0$ and $\cdist(A^{(j)})\ge\gamma$.
\end{enumerate}
\end{corollary}

\begin{proof}
If $\packmod(\alpha)=0$, choose inputs with violation density less than $1/j$. Each has a violation, so $\viol(A^{(j)})\ge\binom{s_j}k^{-1}$, forcing $s_j\to\infty$.

If completion stability fails, choose $\gamma>0$ and $\theta_j\downarrow0$ with $\compmod(\theta_j)>2\gamma$, then choose partial matrices of hole density at most $\theta_j$ and completion distance greater than $\gamma$. A bounded subsequence of their orders would eventually have no holes, a contradiction.

Conversely, either sequence makes the corresponding structural condition fail. Apply Theorem~\ref{thm:structure}.
\end{proof}

\section{Packing amplification}\label{sec:packing}

We determine which metric and geometric families satisfy packing amplification. We first establish positive packing bounds for ultrametrics (Lemma~\ref{lem:ult-packing}) and tree pseudometrics (Lemma~\ref{lem:tree-packing}), reducing the tree case to the ultrametric setting via rooted products. Next, we develop an affine anchor repair method for $d$-dimensional Euclidean pseudometrics (\Cref{prop:euc-pseudo-distance-forces-edits}), yielding both packing amplification (\Cref{lem:packing-euclidean-pseudo}) and quantitative removal bounds (\Cref{cor:euc-removal}). After confirming that packing amplification passes to edgewise restrictions, and thus to strict metric families (\Cref{cor:strict-packing}), we construct a robust metric face (\Cref{lem:metric-face}) to prove that general metrics and metrics with bounded aspect ratio in $[1,C]$ with $C>2$ fail packing amplification and admit no constant-query one-sided testers (\Cref{thm:metric-packing-obstruction}).

\subsection{Ultrametrics and tree pseudometrics}

\begin{lemma}[Packing amplification for ultrametrics]\label{lem:ult-packing}
For every $\alpha>0$,
$$
\packmod_{\mathrm{ult}}(\alpha)
\ge
\min\left\{1,\frac{3}{50}\alpha^2\right\}.
$$
\end{lemma}

\begin{proof}
Let $M\in\Dcal_n$, and let $\mathcal T$ be an edge-disjoint family of $m=\nu(M)$ violating triangles. We prove $|\Vcal(M)|\ge \frac{m^2}{100n}$.

Consider distinct $T,T'\in\mathcal T$ sharing a vertex $v$. By edge-disjointness, they share only $v$. Orient the pair so that the unique maximum in $T$ is at least the unique maximum in $T'$. Write $T=\{a,b,c\}$, with unique maximum edge $ab$ of value $R$. There is $w\in T'\setminus\{v\}$ with $M_{vw}<R$, since one edge of $T'$ incident with $v$ is not its unique maximum.

We claim that $T\cup T'$ contains a violating triangle $Q\ne T,T'$ containing an edge of $T$ and the vertex $w$. If $v=c$, either $\{a,v,w\}$ or $\{b,v,w\}$ violates; otherwise both $M_{aw}$ and $M_{bw}$ are less than $R$, so $\{a,b,w\}$ violates. If $v=a$, either $\{c,v,w\}$ or $\{c,b,w\}$ violates; otherwise the first gives $M_{cw}<R$, the second gives $M_{bw}<R$, and $\{v,b,w\}$ violates. The case $v=b$ is symmetric.

Let $r_x=|\{T\in\mathcal T:x\in T\}|$. Using $\sum_xr_x=3m$ and Cauchy--Schwarz, the number $I$ of unordered intersecting pairs is $I=\sum_x\binom{r_x}{2} \ge \frac12(\frac{9m^2}{n}-3m)$.

If $m<n$, then $|\Vcal(M)|\ge m\ge m^2/n$. If $m\ge n$, then $I\ge3m^2/n$. Choose one mixed violation $Q$ for every intersecting pair. A fixed $Q$ is selected at most $27$ times: choose its edge belonging to $T$ in at most three ways (which determines $T$), choose $v\in T$ in at most three ways, and choose $w\in Q$ in at most three ways (where the edge $vw$ determines $T'$ by edge-disjointness). Hence $|\Vcal(M)|\ge I/27 \ge \frac{m^2}{9n}$ when $m\ge n$. The claimed bound $|\Vcal(M)|\ge\frac{m^2}{100n}$ follows in all cases.

If $m\ge\alpha n^2$, then $\viol(M)\ge \frac{\alpha^2n^3/100}{\binom n3} \ge\frac{3}{50}\alpha^2$.
\end{proof}

To extend this argument to tree pseudometrics, we use the classical correspondence between tree metrics and ultrametrics via the rooted Gromov product (or Farris transform)~\cite{Farris,Gromov}.

\begin{lemma}\label{lem:tree-rooted-char}
Let $n\ge4$ and fix $o\in[n]$. For $i\ne o$, put $h_i=D_{oi}$, and for distinct $i,j\ne o$, put $s_{ij}=(h_i+h_j-D_{ij})/2$. Then $D\in\mathcal T_n$ if and only if $0\le s_{ij}\le\min\{h_i,h_j\}$ for every such pair, and the minimum of $s_{ij},s_{ik},s_{jk}$ is attained at least twice for every triple of distinct labels outside $o$.
\end{lemma}

\begin{proof}
The triangle inequalities involving $o$ are equivalent to the bounds $0\le s_{ij}\le\min\{h_i,h_j\}$. The four-point condition on $\{o,i,j,k\}$ is equivalent to the minimum condition, since $D_{ij}+D_{ok}=h_i+h_j+h_k-2s_{ij}$.

Conversely, the bounds give all triangle inequalities involving $o$. For a triangle outside $o$, we have $D_{ij}+D_{ik}-D_{jk} = 2h_i+2(s_{jk}-s_{ij}-s_{ik}) \ge 2(h_i-\max\{s_{ij},s_{ik}\}) \ge0$.

Choose $C\ge\max_{i\ne j}s_{ij}$ and set $u_{ij}=C-s_{ij}$ off the diagonal and $u_{ii}=0$. The minimum condition makes $u$ a nonnegative ultrametric, which satisfies the four-point condition. To see this directly on four labels, let $R$ be their maximum distance and partition them by the relation $u_{ij}<R$. For partition types $2+2$, $2+1+1$, or $1+1+1+1$, two opposite-edge sums equal $2R$ and the third is at most $2R$. For type $3+1$, the three opposite-edge sums are $R$ plus the three distances of an ultrametric triple, whose maximum is attained at least twice. The all-zero case is immediate.

For a quartet outside $o$, the opposite-edge sums of $D$ are obtained from those of $u$ by multiplication by $2$ and addition of the common constant $h_i+h_j+h_k+h_l-4C$. Thus $D$ satisfies the four-point condition on these quartets as well. The triangle inequalities handle repeated indices, so Buneman's characterization applies.
\end{proof}

\begin{lemma}\label{lem:tree-rooted-packing}
Let $X$ have size $N\ge3$, let $h_i\ge0$, and let $s_{ij}$ be a symmetric assignment on distinct pairs of $X$. Call a triple \emph{rooted-bad} if some pair violates $0\le s_{ij}\le\min\{h_i,h_j\}$, or if its minimum $s$-value is attained uniquely. If an edge-disjoint family contains at least $\beta N^2$ rooted-bad triples, where $\beta>0$, then the density of rooted-bad triples is at least $\min\{1,3\beta^2/200\}$.
\end{lemma}

\begin{proof}
Let the family have size $m\ge\beta N^2$. If at least $m/2$ members contain a bounds violation, choose one such edge from each. These edges are distinct, and counting their incidences with triples gives at least $(m/2)(N-2)/3$ rooted-bad triples, of density at least $\frac{m}{N(N-1)}\ge\beta$.

Otherwise, at least $m/2$ members satisfy the bounds and have a unique minimum. Choose $C\ge\max_{i\ne j}s_{ij}$ and put $u_{ij}=C-s_{ij}$. These triples form an edge-disjoint family of ultrametric violations. Lemma~\ref{lem:ult-packing}, in its counting form, gives at least $\frac{(m/2)^2}{100N}$ such violations, of density at least $3\beta^2/200$. Since a feasible edge-disjoint triple packing has $\beta\le1/6$, the first case also satisfies the stated bound.
\end{proof}

\begin{lemma}[Packing amplification for tree pseudometrics]\label{lem:tree-packing}
For every $\alpha>0$,
$$
\packmod_{\mathrm{tree}}(\alpha)
\ge
\min\left\{\frac1{1820},\frac{3\alpha^2}{1600}\right\}.
$$
\end{lemma}

\begin{proof}
Let $M\in\Dcal_n$, and let $\mathcal Q$ be an edge-disjoint family of $m\ge\alpha n^2$ violating quartets. For a root $o$, let $r_o=|\{Q\in\mathcal Q:o\in Q\}|$. Since $\sum_or_o=4m$, at most seven roots have $r_o>m/2$. For $n\ge16$, at least $n/2$ roots are good, meaning $r_o\le m/2$.

Fix a good root $o$ and form its rooted data. By Lemma~\ref{lem:tree-rooted-char}, every $Q\in\mathcal Q$ not containing $o$ contains a rooted-bad triple. Choose one such triple from each $Q$. These triples are edge-disjoint and number at least $m/2$, so their packing density on $N=n-1$ vertices is at least $\alpha/2$. Lemma~\ref{lem:tree-rooted-packing} gives a rooted-bad triple density of at least $\min\{1,3\alpha^2/800\}$. Each rooted-bad triple, together with $o$, forms a violating quartet. Summing over good roots and dividing by the maximum multiplicity four gives $\viol(M) \ge \frac{(n/8)\binom{n-1}{3}}{\binom n4} \min\{1,3\alpha^2/800\} = \min\{\frac12,3\alpha^2/1600\}$. For $4\le n<16$, any violation gives density at least $1/1820$. Taking the infimum completes the proof.
\end{proof}

\subsection{Euclidean pseudometrics and edgewise restrictions}

To bound the distance to $\mathcal E_n^{(d)}$ in terms of $(d+3)$-point violations, we adapt the affine anchor repair method of Parnas and Ron~\cite[Section~6]{PR} (see also Krauthgamer and Sasson~\cite{KS} and Onak~\cite{Onak}). In their analysis~\cite[Fact~2 and Lemma~6.1]{PR}, given an anchor set $U$ realized with affine dimension $s\le d$, vertices whose distances to $U$ admit a realization in the same $s$-dimensional affine span (\emph{strongly consistent} points) have uniquely determined coordinates. A global Euclidean pseudometric is then obtained by placing those vertices at their forced coordinates and collapsing the remaining exceptional vertices. Whereas~\cite{PR,KS} locate a suitable dimension $s\le d$ via adaptive or iterative phases that increment the dimension, we select a dimension level $s$ nonadaptively via a pigeonhole threshold on the $(d+3)$-set dimension densities.

\begin{proposition}[Euclidean repair and hitting-set bound]\label{prop:euc-pseudo-distance-forces-edits}
Let $n\ge2d+8$ and $0<\viol(D)< \min\{\frac12,(2d+2)^{-(d+2)}\}$ with respect to $\Phi_{\mathrm{Euc}}^{(d)}$. Put $C_d=3(d+1)\binom{d+5}{2}$. There exists $B\in\mathcal E_n^{(d)}$ differing from $D$ on at most $\frac12 C_dn^2\viol(D)^{1/(d+2)}$ edges. In particular,
$$
\dist_n(D,\mathcal E_n^{(d)})
\le C_d\viol(D)^{1/(d+2)}
\qquad\text{and}\qquad
\frac{2\nu(D)}{n^2}
\le C_d\viol(D)^{1/(d+2)}.
$$
\end{proposition}

\begin{proof}
Let $\mathcal C_j$ be the family of valid $(d+3)$-sets of affine dimension $j$, and put $p_j=|\mathcal C_j|/\binom n{d+3}$, so that $\sum_{j=0}^dp_j=1-\viol(D)$. By averaging, some $p_j$ satisfies $p_j\ge\frac{1-\viol(D)}{d+1} > \frac1{2d+2} > \viol(D)^{1/(d+2)} \ge \viol(D)^{(j+1)/(d+2)}$. Choose $s=\max\{j:p_j\ge\viol(D)^{(j+1)/(d+2)}\}$ and let $\mathcal F=\Vcal(D)\cup\bigcup_{j>s}\mathcal C_j$, which has size $|\mathcal F|=\binom n{d+3}(\viol(D)+\sum_{j>s}p_j)$. By the maximality of $s$ and the lower bound on $p_s$,
$$
|\mathcal F|
\le
\binom n{d+3}(d-s+1)\viol(D)^{(s+2)/(d+2)}
\le
p_s\binom n{d+3}(d+1)\viol(D)^{1/(d+2)}.
$$

Choose an ordered $(d+5)$-tuple of distinct vertices $(c_1,\ldots,c_{d+3},u,v)$ uniformly, write $C=\{c_1,\ldots,c_{d+3}\}$, and condition on $C\in\mathcal C_s$. Every fixed $(d+3)$-set of positions is unconditionally uniform among all $(d+3)$-sets, so its conditional probability of belonging to $\mathcal F$ is at most $|\mathcal F|/(p_s\binom n{d+3})$. A union bound and averaging give a fixed $C\in\mathcal C_s$ for which the number $b$ of \emph{bad} pairs $\{u,v\}\subseteq[n]\setminus C$ satisfying $\binom{C\cup\{u,v\}}{d+3}\cap\mathcal F\ne\varnothing$ is at most $b \le \frac{n^2}{2} \binom{d+5}{d+3} |\mathcal F|/(p_s\binom n{d+3})$. Call the remaining pairs in $[n]\setminus C$ \emph{good}.

Fix a realization $z_c$ of $D[C]$, let $A$ be its $s$-dimensional affine span, and choose an affine basis from $C$. If $\{u,v\}$ is good, every $(d+3)$-subset of $C\cup\{u,v\}$ is realizable, so Lemma~\ref{lem:euclidean-local-char} gives a realization of $C\cup\{u,v\}$. After aligning $C$, both new points must lie in $A$; otherwise an affinely independent subset of dimension greater than $s$, padded to a $(d+3)$-set, would belong to $\mathcal F$. By Lemma~\ref{lem:affine-basis-unique}, their positions in $A$ are uniquely determined by the affine basis.

Let $X$ be the set of vertices outside $C$ incident to at least one good pair. Each $u\in X$ therefore has a well-defined coordinate $x_u\in A$ agreeing with all its prescribed distances to $C$, and every good pair satisfies $D_{uv}=\|x_u-x_v\|_2$. Let $I=([n]\setminus C)\setminus X$. Every pair incident to $I$ inside $[n]\setminus C$ is bad, so $|I|(n-d-4)\le2b$. Since $n\ge2d+8$, this implies $|I|\le4b/n$.

Assign the points of $C$ their fixed realization, assign $u\in X$ to $x_u$, and collapse every vertex of $I$ to a single fixed basis point. The resulting Euclidean pseudometric $B$ differs from $D$ only on bad pairs outside $C$ and on edges between $C$ and $I$, so $|\{e:B_e\ne D_e\}| \le b+(d+3)|I| \le 3b \le \frac12 C_dn^2\viol(D)^{1/(d+2)}$. The normalized repair bound follows. Since every original violation contains a changed edge, edge-disjoint violations require distinct changed edges, yielding the packing bound.
\end{proof}

\begin{lemma}[Packing amplification for Euclidean pseudometrics]\label{lem:packing-euclidean-pseudo}
Put $C_d=3(d+1)\binom{d+5}{2}$ and $\delta_d = \min\{\frac12, (2d+2)^{-(d+2)}, \binom{2d+7}{d+3}^{-1}\}$. For every $\alpha>0$,
$$
\packmod_{\mathrm{Euc}}^{(d)}(\alpha)
\ge
\min\left\{\delta_d,\left(\frac{2\alpha}{C_d}\right)^{d+2}\right\}.
$$
\end{lemma}

\begin{proof}
Suppose $\nu(D)\ge\alpha n^2$. If $\viol(D)\ge\delta_d$, there is nothing to prove. Otherwise, positivity of $\nu(D)$ gives $\viol(D)\ge\binom n{d+3}^{-1}$, and the definition of $\delta_d$ forces $n\ge2d+8$. Proposition~\ref{prop:euc-pseudo-distance-forces-edits} then gives $2\alpha\le C_d\viol(D)^{1/(d+2)}$. Taking the infimum completes the proof.
\end{proof}

\begin{corollary}[Quantitative Euclidean removal]\label{cor:euc-removal}
For every $\epsilon>0$, with $C_d$ and $\delta_d$ as in Lemma~\ref{lem:packing-euclidean-pseudo}, we have $\rho_{\mathrm{Euc}}^{(d)}(\epsilon) \ge \min\{\delta_d,(\epsilon/C_d)^{d+2}\}$.
\end{corollary}

\begin{proof}
An $\epsilon$-far input has positive violation density. If $\viol(D)<\delta_d$, then $n\ge 2d+8$ as in Lemma~\ref{lem:packing-euclidean-pseudo}, and Proposition~\ref{prop:euc-pseudo-distance-forces-edits} implies $\epsilon<C_d\viol(D)^{1/(d+2)}$. Taking the infimum proves the claim.
\end{proof}

\begin{proposition}\label{prop:pseudometric-to-metric-packing}
Let $\Gamma\subseteq\Lambda$ and $\Phi'=\Phi\cap\Gamma^{\binom{[k]}2}$, regarding both predicates on the ambient value set $\Lambda$. Then for every $\alpha>0$,
$$
\packmod_{\Phi'}(\alpha)
\ge
\sup_{0<t<1}
\min\left\{
2(1-t)\alpha,\,
\packmod_\Phi(t\alpha)
\right\}
\ge
\min\left\{\alpha,\packmod_\Phi(\alpha/2)\right\}.
$$
Thus packing amplification passes to edgewise restrictions. The assertion concerns violation sets and remains meaningful even when a restricted valid family is empty at some orders.
\end{proposition}

\begin{proof}
Fix an edge-disjoint family $\mathcal S$ of $\Phi'$-violations with $|\mathcal S|\ge\alpha n^2$. For any $t\in(0,1)$, either at least $t\alpha n^2$ members violate $\Phi$, or at least $(1-t)\alpha n^2$ members satisfy $\Phi$ but contain a value outside $\Gamma$.

In the first case, $\viol_{\Phi'}(D)\ge\viol_\Phi(D)\ge\packmod_\Phi(t\alpha)$. In the second, edge-disjointness gives a set $Z$ of at least $(1-t)\alpha n^2$ distinct edges with values outside $\Gamma$. Counting incidences gives $|\Vcal_{\Phi'}(D)|\binom k2 \ge |Z|\binom{n-2}{k-2}$, and hence $\viol_{\Phi'}(D)\ge\frac{2|Z|}{n(n-1)} \ge2(1-t)\alpha$. Taking the infimum over $D$ and the supremum over $t$ yields the first inequality; setting $t=1/2$ gives the second.
\end{proof}

\begin{corollary}[Packing amplification for strict geometric families]\label{cor:strict-packing}
On $\Lambda=\mathbb R_{\ge0}$ or $\Lambda=(0,\infty)$, the strict predicates $\Phi_{\mathrm{ult}}^+$, $\Phi_{\mathrm{tree}}^+$, and $\Phi_{\mathrm{Euc}}^{(d),+}$ all satisfy packing amplification.
\end{corollary}

\begin{proof}
Apply Proposition~\ref{prop:pseudometric-to-metric-packing} with $\Gamma=(0,\infty)$ to Lemmas~\ref{lem:ult-packing}, \ref{lem:tree-packing}, and~\ref{lem:packing-euclidean-pseudo}. Restricting the infimum to positive inputs can only increase the modulus.
\end{proof}

\subsection{Robust metric faces and the packing obstruction for general metrics}

Unlike the geometric families above, general triangle inequalities admit dense edge-disjoint families of tight triangles with every other slack strictly positive. In contrast to tripartite graph constructions for triangle-freeness (such as the Behrend graph construction of~\cite[Section~5.1]{AKKR}, which also underlies the metric testing lower bound in~\cite[Section~4]{BKW}) that restrict edge steps to a progression-free set and incur an $n^{-o(1)}$ density loss, a metric face on a complete graph only requires unselected triangles to have strictly positive slack. The quadratic potential $\lambda(i+j-k)^2$ in Lemma~\ref{lem:metric-face} below (and its midpoint variant in Lemma~\ref{lem:metric-blowup}) exploits this real-valued margin to achieve full $\Theta(n^2)$ edge-disjoint packing density without progression-free sets; Section~\ref{sec:quantitative-testing} later combines these same faces with signed noise.

\begin{lemma}[Robust metric face]\label{lem:metric-face}
Fix $C>2$. There are a rational $C_0$ with $2<C_0<\min\{C,3\}$ and an absolute constant $\beta=1/20$ such that, for all sufficiently large $n$, there exist a rational metric $A\in[1,C_0]^{\binom{[n]}2}$ and an edge-disjoint family $\mathcal H$ of at least $\beta n^2$ triangles satisfying:
\begin{enumerate}[label=(\arabic*)]
\item each member of $\mathcal H$ has one designated tight triangle inequality, and these are the only tight triangle inequalities of $A$;
\item every other triangle inequality has slack at least $n^{-3}$, and every entry has distance at least $n^{-3}$ from the endpoints of $[1,C_0]$;
\item every entry has an $O_C(\log n)$-bit rational representation.
\end{enumerate}
\end{lemma}

\begin{proof}
Choose a positive rational $\rho<\min\{(C-2)/10,1/100\}$, and set $C_0=2+10\rho$ and $b=1+4\rho$. First consider $4m-1$ vertices partitioned into $X=\{x_i:|i|\le m-1\}$, $Y=\{y_j:j\in[m]\}$, and $Z=\{z_k:k\in[m]\}$. Put $\lambda=\rho/m^2$, assign value $3/2$ within each part, and set
$$
A(x_i,y_j)=b+\lambda(i^2+2ij),\qquad
A(y_j,z_k)=b+\lambda(j^2-2jk),\qquad
A(x_i,z_k)=2b+\lambda(2ik-k^2).
$$
Since $2ik-k^2=i^2-(i-k)^2<m^2$, the index ranges imply $A(x_i,y_j)\in[1+3\rho,1+7\rho]$, $A(y_j,z_k)\in[1+3\rho,1+5\rho]$, and $A(x_i,z_k)\in[2+5\rho,2+9\rho]$.

For a triangle meeting all three parts, the slack on $(x_i,z_k)$ is
$$
A(x_i,y_j)+A(y_j,z_k)-A(x_i,z_k)
=
\lambda(i+j-k)^2,
$$
while the other two slacks are at least $2+\rho$. If two vertices lie in one part, their edge has value $3/2$ and the two cross-part entries differ by at most $4\rho$, giving slacks at least $3/2-4\rho$ and $2(1+3\rho)-3/2$. Triangles within a part are equilateral and hence strict.

Thus the tight triangles are precisely $\mathcal H = \{\{x_{k-j},y_j,z_k\}:j,k\in[m]\}$, with designated long edge $x_{k-j}z_k$. Any two vertices in different parts determine at most one third vertex in $\mathcal H$, so $\mathcal H$ is edge-disjoint and has size $m^2$.

For arbitrary sufficiently large $n$, take $m=\lfloor(n+1)/4\rfloor$ and add the at most three remaining vertices at distance $3/2$ from every other vertex; any triangle meeting a new vertex and an old edge of value $a$ has positive slacks $a,a,3-a$. The strict slacks are at least $\min\{\rho/m^2,1/2,3-C_0\}$ and the endpoint margins are at least $\rho$, both exceeding $n^{-3}$ for large $n$. Finally, $m^2/n^2\to1/16>\beta$, and the entries have logarithmic bit complexity.
\end{proof}

\begin{theorem}[Packing obstruction for general metrics]\label{thm:metric-packing-obstruction}
For $\Lambda=[1,C]$ with $C>2$, the metric predicate admits a packing obstruction: for $\beta=1/20$,
$$
\packmod(\beta)=0
\qquad\text{and}\qquad
\rho(\epsilon)=0
\quad(0<\epsilon<2\beta).
$$
The same holds for positive-valued general metrics on $(0,\infty)$ and nonnegative-valued general pseudometrics on $[0,\infty)$. Consequently, none of these families has a constant-query one-sided tester.
\end{theorem}

\begin{proof}
Take $(A,\mathcal H)$ from Lemma~\ref{lem:metric-face}, with $h=|\mathcal H|\ge\beta n^2$, and add $\eta=n^{-4}$ to the designated long edge of each member of $\mathcal H$. Each selected triangle now has designated slack $-\eta$. Every formerly strict slack changes by at most $3\eta<n^{-3}$, and the endpoint margins keep all entries in $[1,C_0]$. Thus the resulting matrix $D$ has $\Vcal(D)=\mathcal H$, which gives $\nu(D)=h\ge\beta n^2$ and $\viol(D)=h/\binom n3=O(n^{-1})\to0$.

Every metric or pseudometric repair changes at least one edge in each member of $\mathcal H$, while resetting the $h$ long edges restores $A$. Hence $\dist_n(D,\mathcal M_n[1,C]) = \dist_n(D,\mathcal M_n) = \dist_n(D,\overline{\mathcal M}_n) = 2h/n^2\ge2\beta$. The removal claims follow immediately, and Theorem~\ref{thm:testing} together with extendability rules out constant-query one-sided testers.
\end{proof}

\begin{remark}[Transfer to $\delta$-hyperbolicity]\label{rem:hyperbolic-packing}
Following Gromov~\cite{Gromov}, a metric is $\delta$-hyperbolic if $d_{ij}+d_{kl} \le \max\{d_{ik}+d_{jl},d_{il}+d_{jk}\}+2\delta$ for all labels; the corresponding four-point predicate includes the triangle inequalities. Fix $\delta>0$ and a positive rational $\lambda\le\delta/C_0$, with $A,D,\mathcal H$ and $h=|\mathcal H|$ as in the proof of Theorem~\ref{thm:metric-packing-obstruction}. Because every entry of $\lambda A$ and $\lambda D$ is at most $\delta$, the four-point hyperbolicity inequalities hold automatically, so the violations of $\lambda D$ are precisely the $h(n-3)$ quartets containing a triangle of $\mathcal H$. Since any edge belongs to at most one triangle in $\mathcal H$ and meets at most $n-1$ others at a single endpoint, each edge belongs to at most $2n$ violating quartets. Greedy selection then yields at least $h(n-3)/(12n)\ge h/24$ edge-disjoint violating quartets for $n\ge6$ while the violation density remains $O(n^{-1})$, giving a packing obstruction for $\delta$-hyperbolic metrics.

Moreover, because scaling preserves exact Hamming discrepancies, every scaled bad input remains at least $2\beta$-far from $\delta$-hyperbolic metrics (with $\lambda A$ as a valid target). Applying the same scaling to the signed noise distributions in \Cref{sec:quantitative-testing} for any fixed $C>2$ transfers our quantitative nonadaptive and adaptive query lower bounds for general metrics to $\delta$-hyperbolic metrics.
\end{remark}

\section{Completion stability}\label{sec:completion}

We establish completion stability for our principal pseudometric and metric families and identify the completion obstructions created by injective geometric realization. We first prove linear completion stability for ultrametrics (\Cref{lem:ult-completion}) and tree pseudometrics (\Cref{lem:tree-completion-stability}) via a recursive graph separation scheme, and establish $O(\sqrt{\theta})$ completion stability for $d$-dimensional Euclidean pseudometrics (\Cref{lem:completion-stability-euclidean-pseudo}) via affine anchor rigidity. We next analyze metrics with bounded aspect ratio on $[1,C]$: after showing that $C=3$ is the exact completion threshold (Proposition~\ref{prop:exact-small-c}), we use a cyclic entropy inequality to prove linear completion stability for every fixed $C>1$ (Theorem~\ref{thm:completion-stability-C}), which in turn yields uniform polynomial-time completion repair (Corollary~\ref{cor:algorithmic-completion-repair}) and a triangle packing bound for metric repair (Proposition~\ref{prop:metric-repair-approx}). Finally, we use a collinear twin construction (Lemma~\ref{lem:twin-rigidity}) to prove that strict tree metrics and strict Euclidean metrics admit completion obstructions and are not constant-query one-sided testable (Theorem~\ref{thm:completion-obstructions}). Combining our results from Sections~\ref{sec:packing} and~\ref{sec:completion} shows that packing amplification and completion stability are logically independent (Corollary~\ref{cor:independence}).

\subsection{Ultrametrics and tree pseudometrics via recursive separation}

Our recursive separation scheme for completing locally consistent partial ultrametrics and tree pseudometrics complements algorithmic work on fitting and repairing ultrametrics and tree metrics under edge disagreements or outlier deletions~\cite{SWW,CFLM,BLLSSSX}. We begin with a structural graph partitioning lemma that bounds the number of $F$-edges cut when separating all $R$-edges in an edge tripartition without forbidden triangles.

\begin{lemma}\label{lem:separation}
Let $F,R,S$ partition $E(V)=\binom V2$, and suppose no triangle has two edges in $F$ and one in $R$. There exist a partition $\mathcal P$ of $V$, a set $Z\subseteq F$, and a set $Q\subseteq S$ such that:
\begin{enumerate}[label=(\arabic*)]
\item no part contains an $R$-edge;
\item $Z$ is the set of $F$-edges crossing $\mathcal P$;
\item every edge of $Q$ crosses $\mathcal P$;
\item $|Z|\le3|Q|$.
\end{enumerate}
\end{lemma}

\begin{proof}
We first verify that if $p_0p_1p_2p_3$ is a shortest path in a graph $G$, then some $p_i$ satisfies $\deg_G(p_i)\le3|N_G^2(p_i)|$, where $N_G^2(x)$ denotes the set of vertices at distance exactly two from $x$. For each vertex $x$, let $A_x=\{i:xp_i\in E(G)\}$ and $B_x=\{i:\dist_G(x,p_i)=2\}$. Because $p_0p_1p_2p_3$ is a shortest path, no vertex is adjacent to both $p_0$ and $p_3$, so $|A_x|\le3$. Moreover, if $A_x\ne\varnothing$, then $B_x\ne\varnothing$: for a path vertex this follows from the shortest-path property, and for any other vertex a transition between an adjacent and a nonadjacent consecutive path vertex yields distance two. Thus $|A_x|\le3|B_x|$, and summing over $x$ proves the claim.

We now work separately in the connected components of $(V,F)$ and induct on $|V|$. If a component contains no internal $R$-edge, take it as a single part. Otherwise, choose $ab\in R$ and a shortest $F$-path from $a$ to $b$. Its length is at least three, as length one is impossible and length two would form a forbidden triangle. By the initial claim, some vertex $v$ among the first four path vertices satisfies $\deg_F(v)\le3|N_F^2(v)|$. Set $Z_0=\{vw\in F:w\in V\}$ and $Q_0=\{vw:w\in N_F^2(v)\}$. Every edge of $Q_0$ lies in $S$, since it is not in $F$ and membership in $R$ would create a forbidden triangle. Apply the induction hypothesis to the $F$-components of $V\setminus\{v\}$, add $\{v\}$ as a singleton part, and adjoin $Z_0$ and $Q_0$ to the recursive sets. Since the new edges are incident to $v$ while the recursive edges are not, the cardinality bounds add and all four properties follow.
\end{proof}

\begin{lemma}[Completion stability for ultrametrics]\label{lem:ult-completion}
Over any nonempty value set $\Lambda$, a locally consistent partial ultrametric with $h$ holes can be completed to a valid ultrametric by changing at most $3h$ specified entries. Consequently,
$$
\compmod_{\mathrm{ult}}(\theta)\le\min\{1,6\theta\}.
$$
This includes $\mathcal U$ on $\mathbb R_{\ge0}$ and $\mathcal U^+$ on $(0,\infty)$.
\end{lemma}

\begin{proof}
Let $A:D\to\Lambda$ be the specified assignment on a finite vertex set $V$, and let $H=E(V)\setminus D$ be its hole set. We recursively process subsets $X\subseteq V$ with a cap $\beta\in\Lambda$ bounding all specified values inside $X$, initialized to the maximum specified value (or any element of $\Lambda$ if $D=\varnothing$).

If no edge inside $X$ is specified, assign $\beta$ to all edges of $E(X)$. Otherwise, let $\lambda_X=\max_{e\in D\cap E(X)}A_e$ and partition $E(X)$ into $F_X=\{e\in D\cap E(X):A_e<\lambda_X\}$, $R_X=\{e\in D\cap E(X):A_e=\lambda_X\}$, and $S_X=H\cap E(X)$. Local consistency rules out triangles with two $F_X$-edges and one $R_X$-edge. Apply Lemma~\ref{lem:separation}, assign $\lambda_X$ to all edges crossing the resulting partition, and recurse on each part with cap $\lambda_X$.

Because $R_X\ne\varnothing$ and no part contains an $R_X$-edge, each recursive part is proper and the recursion terminates. Any triangle crossing a partition has at least two edges of value $\lambda_X$ and its third edge at most $\lambda_X$, while triangles within a part are valid by induction; hence the output is a valid ultrametric in $\Lambda$. Among specified edges, only the crossing $F_X$-edges $Z_X$ are modified, and Lemma~\ref{lem:separation} gives $|Z_X|\le3|Q_X|$ for a subset $Q_X\subseteq H$ crossing the partition. The sets $Q_X$ are pairwise disjoint across the recursion tree because every edge of $Q_X$ crosses the partition of $X$, whereas descendants lie within single parts and incomparable nodes occupy disjoint vertex sets. Thus at most $3|H|$ specified edges are modified in total, yielding $\cdist(A)\le6|H|/n^2$ for $n\ge3$.
\end{proof}

\begin{corollary}\label{cor:strict-ult-completion}
On the ambient value set $\mathbb R_{\ge0}$, we have $\compmod_{\mathrm{ult}^+}(\theta)\le\min\{1,10\theta\}$. On the positive ambient value set $(0,\infty)$, the bound improves to $\min\{1,6\theta\}$.
\end{corollary}

\begin{proof}
The bound on $(0,\infty)$ follows from \Cref{lem:ult-completion}. On $\mathbb R_{\ge0}$, let $A$ be locally consistent for $\Phi_{\mathrm{ult}}^+$ with $h$ holes and $z$ specified zero edges. Every triple containing a specified zero edge must contain a hole, so in each triple the number of specified zero edges is at most twice the number of holes. Summing over all triples gives $z(n-2)\le2h(n-2)$, so $z\le2h$. Replacing all specified zeros with a positive value leaves every fully specified triple unchanged; applying Lemma~\ref{lem:ult-completion} over $(0,\infty)$ then changes at most $3h$ additional specified entries, for at most $z+3h\le5h$ changes in total.
\end{proof}

To extend Lemma~\ref{lem:ult-completion} to tree pseudometrics, we pass to rooted Gromov products as in Lemma~\ref{lem:tree-rooted-char}.

\begin{lemma}\label{lem:tree-rooted-completion}
Let $X$ be a finite set with heights $h_i\ge0$, and let $s$ be a partial symmetric assignment on $D\subseteq\binom X2$ with hole set $H=\binom X2\setminus D$. Suppose $0\le s_{ij}\le\min\{h_i,h_j\}$ for all $ij\in D$, and every fully specified triple has its minimum $s$-value attained at least twice. Then there exists a full assignment $\widetilde s$ satisfying both conditions everywhere and differing from $s$ on at most $3|H|$ specified entries.
\end{lemma}

\begin{proof}
Assume $X\ne\varnothing$, and set $H_0=\max\bigl(\{0\}\cup\{h_i:i\in X\}\bigr)$ and $u_{ij}=H_0-s_{ij}$ for $ij\in D$. The specified $u$-values lie in $[0,H_0]$ and are locally ultrametric, so Lemma~\ref{lem:ult-completion} produces a full ultrametric $U$ in $[0,H_0]$ differing from $u$ on at most $3|H|$ specified entries. For distinct $i,j\in X$, put $v_{ij}=H_0-U_{ij}$ and $\widetilde s_{ij}=\min\{v_{ij},h_i,h_j\}$. Then $\widetilde s_{ij}$ is nonnegative, satisfies the height bounds, and inherits the minimum condition from $v_{ij}\ge\min\{v_{ik},v_{jk}\}$ via
$\min\{\widetilde s_{ik},\widetilde s_{jk}\} = \min\{v_{ik},v_{jk},h_i,h_j,h_k\} \le \min\{v_{ij},h_i,h_j\} = \widetilde s_{ij}$.
Whenever $U_{ij}=u_{ij}$ on a specified pair, we have $v_{ij}=s_{ij}\le\min\{h_i,h_j\}$ and hence $\widetilde s_{ij}=s_{ij}$, introducing no further discrepancies.
\end{proof}

\begin{lemma}[Completion stability for tree pseudometrics]\label{lem:tree-completion-stability}
For every $\theta\ge0$,
$$
\compmod_{\mathrm{tree}}(\theta)\le\min\{1,26\theta\}.
$$
\end{lemma}

\begin{proof}
Let $A$ be locally consistent on $[n]$ with $n\ge4$ and $h=|H(A)|$. Choose a root $o$ of minimum hole degree $y=\deg_H(o)\le 2h/n$, and set $X=\{i\ne o:oi\notin H(A)\}$. If $N=|X|\le2$, then $y\ge n-3$, so $h\ge n(n-3)/2$ and $\cdist(A)\le1\le26h/n^2$. Assume henceforth that $N\ge3$.

Let $H_X=H(A)\cap\binom X2$ and $h_X=|H_X|$. For specified pairs in $X$, set $h_i=A_{oi}$ and $s_{ij}=(h_i+h_j-A_{ij})/2$, and let $E_0$ be the set of specified pairs in $X$ violating $0\le s_{ij}\le\min\{h_i,h_j\}$, with $b=|E_0|$. Any triple $T\in\binom X3$ meeting $E_0$ must contain a hole in $H_X$, as otherwise $T\cup\{o\}$ would be a fully specified quartet violating a triangle inequality. Thus $|E_0\cap\binom T2|\le 2|H_X\cap\binom T2|$ for every $T\in\binom X3$; summing over $T$ and dividing by $N-2$ gives $b\le2h_X$.

Deleting $E_0$ leaves a partial assignment on $X$ with $h_X+b$ holes that satisfies the hypotheses of Lemma~\ref{lem:tree-rooted-completion}, since every remaining fully specified triple forms a valid quartet with $o$. Lemma~\ref{lem:tree-rooted-completion} yields a completion $\widetilde s$ on $X$ changing at most $3(h_X+b)$ remaining specified entries, so at most $b+3(h_X+b)\le11h_X$ originally specified entries in $X$ are modified. Setting $B_{oi}=h_i$ and $B_{ij}=h_i+h_j-2\widetilde s_{ij}$ for distinct $i,j\in X$ defines a tree pseudometric on $\{o\}\cup X$ by Lemma~\ref{lem:tree-rooted-char}. Identifying each of the $y$ vertices outside $\{o\}\cup X$ with $o$ alters at most $yn\le2h$ additional specified entries, for a total of at most $11h_X+2h\le13h$ modified specified entries and $\cdist(A)\le26h/n^2$.
\end{proof}

\begin{corollary}\label{cor:tree-completion-sharp}
For every $\theta>0$, we have $\frac{\theta}{1+8\theta}\le\compmod_{\mathrm{tree}}(\theta)\le\min\{1,26\theta\}$, and in particular $\compmod_{\mathrm{tree}}(\theta)=\Theta(\theta)$ as $\theta\downarrow0$.
\end{corollary}

\begin{proof}
For the lower bound, take $m=\lceil 1/(8\theta)\rceil$ disjoint four-vertex gadgets on $n=4m$ vertices. In each gadget, specify a four-cycle with consecutive edge weights $1,1,1,4$ and leave both diagonals as holes; set all cross-gadget distances to $4$. Since every triple inside a gadget contains a hole, any fully specified quartet meets each gadget in at most two vertices, so two of its opposite-edge sums equal $8$, the third is at most $8$, and all triangle inequalities hold. Thus the partial matrix is locally consistent for $\Phi_{\mathrm{tree}}$, yet every metric repair must edit at least one cycle edge per gadget because $4>1+1+1$. This gives $|H(A)|/n^2=1/(8m)$ and $\cdist(A)\ge1/(8m)$, matching the upper bound of Lemma~\ref{lem:tree-completion-stability}.
\end{proof}

\subsection{Euclidean pseudometrics via affine anchor rigidity}

For $d$-dimensional Euclidean pseudometrics, we adapt the affine anchor repair strategy of~\cite[Lemma~6.1]{PR} to partial matrices by selecting a fully specified $(d+3)$-clique of maximum affine dimension among vertices of low hole degree.

\begin{lemma}[Completion stability for Euclidean pseudometrics]\label{lem:completion-stability-euclidean-pseudo}
For every $\theta\ge0$,
$$
\compmod_{\mathrm{Euc}}^{(d)}(\theta)
\le
\min\left\{1,2(d+5)\sqrt{\theta}\right\}.
$$
\end{lemma}

\begin{proof}
Let $A$ be locally consistent on $[n]$ with $n\ge d+3$. The bound is trivial if $H(A)=\varnothing$ or $\theta\ge 1/(4(d+5)^2)$, so assume $0<\theta<1/(4(d+5)^2)$ and $1\le |H(A)|\le\theta n^2$; in particular, $n\ge1/\sqrt{\theta}$. Let $G$ be the graph of specified pairs and $W=\{v\in[n]:\deg_H(v)\le\sqrt{\theta}n\}$, so $|[n]\setminus W|\le2\sqrt{\theta}n$ by the handshake lemma. A greedy selection finds a $(d+3)$-clique in $G[W]$: after choosing $j\le d+2$ vertices, the number of common neighbors in $W$ is at least $n(1-(j+2)\sqrt{\theta})-j \ge \theta^{-1/2}-2(d+3)>4$.

Choose a $(d+3)$-clique $C$ in $G[W]$ whose realization has maximum affine dimension $s$, fix a realization in $\mathbb R^s$ with affine basis $c_0,\ldots,c_s\in C$, and set $X=\{v\in W:vc\notin H(A)\text{ for all }c\in C\setminus\{v\}\}$. Then $C\subseteq X$ and $|[n]\setminus X| \le 2\sqrt{\theta}n+\sum_{c\in C}\deg_H(c) \le (d+5)\sqrt{\theta}n$. For each $v\in X\setminus C$, the submatrix on $C\cup\{v\}$ is fully specified and locally valid, so Lemma~\ref{lem:euclidean-local-char} yields a realization in $\mathbb R^d$. Its affine dimension cannot exceed $s$, as otherwise padding an affinely independent $(s+2)$-subset with vertices of $C$ would contradict the maximality of $s$. After aligning $C$, the image of $v$ lies in $\mathbb R^s$ and is uniquely determined by Lemma~\ref{lem:affine-basis-unique}; denote it by $p_v$.

For any specified pair $uv$ in $X$, padding $\{u,v,c_0,\ldots,c_s\}$ with vertices of $C$ forms a fully specified $(d+3)$-set in $W$ of affine dimension at most $s$, and aligning the affine basis forces $A_{uv}=\|p_u-p_v\|_2$. Assigning every vertex outside $X$ to $p_{c_0}$ produces a global Euclidean pseudometric agreeing with $A$ on all specified pairs inside $X$, so $\cdist(A)\le \frac{2}{n^2}|[n]\setminus X|n \le 2(d+5)\sqrt{\theta}$.
\end{proof}

\subsection{Metrics with bounded aspect ratio via cyclic entropy}

For a partial matrix $A$ with specified values in $[1,C]$, let $G_A$ denote its graph of specified edges. A simple cycle in $G_A$ is \emph{nonmetric} (also called \emph{broken} or \emph{unbalanced} in the metric repair literature~\cite{BDST,GJ,FRV,FGRSV}) if one edge weight strictly exceeds the sum of the remaining edge weights. We first record the bounded range form of the classical metric completion criterion~\cite{HY,DL}.

\begin{lemma}\label{lem:metric-completion-criterion}
A partial matrix with specified values in $[1,C]$ has a metric completion in $[1,C]$ preserving all specified entries if and only if $G_A$ contains no nonmetric cycle.
\end{lemma}

\begin{proof}
Necessity is immediate from the triangle inequality along any cycle. Conversely, let $d_G$ denote shortest-path distance in $G_A$ (with $d_G(u,v)=+\infty$ between components), and set $B_{uv}=\min\{C,d_G(u,v)\}$ for $u\ne v$ and $B_{uu}=0$. If $uv$ is specified, a strict inequality $d_G(u,v)<A_{uv}$ would yield a shorter simple path in $G_A$ and hence a nonmetric cycle, so $B$ preserves all specified entries. Moreover, every path between distinct vertices has length at least $1$, and for any distinct $u,v,w$, either $\max\{d_G(u,w),d_G(w,v)\}\ge C$ gives $B_{uw}+B_{wv}\ge C\ge B_{uv}$, or else $B_{uv}\le d_G(u,w)+d_G(w,v)=B_{uw}+B_{wv}$. Thus $B\in\mathcal M_n[1,C]$.
\end{proof}

The number of edits to specified edges in $\cdist(A)$ is the sparse metric repair objective on $G_A$ normalized by $2/n^2$, or equivalently the minimum size of a hitting set of specified edges for all nonmetric cycles~\cite{GJ,FRV,FGRSV}. We now bound this optimum in terms of the hole count under local consistency, beginning with the exact completion threshold at $C=3$ (which differs from the trivial threshold $C=2$ below which every complete input is already metric).

\begin{proposition}[Exact completion threshold at $C=3$]\label{prop:exact-small-c}
If $1<C\le3$, every locally consistent partial matrix in $[1,C]$ has an exact metric completion, so $\compmod(\theta)=0$ for all $\theta\ge0$. Conversely, for every $C>3$, locally consistent partial matrices without exact completions exist, and $\compmod(\theta)\ge\frac{\theta}{1+8\theta}$ for all $\theta>0$.
\end{proposition}

\begin{proof}
Local consistency rules out nonmetric triangles, and when $C\le3$ any cycle of length $r\ge4$ has maximum edge weight at most $3\le r-1$, excluding longer nonmetric cycles; Lemma~\ref{lem:metric-completion-criterion} then gives an exact completion. For $C>3$, fix $L\in(3,C]$ and apply the $m$-gadget four-cycle construction of Corollary~\ref{cor:tree-completion-sharp} with cycle weights $1,1,1,L$, diagonal holes, and cross-gadget distances $C$. Every fully specified triangle is valid, each gadget requires at least one specified edit, and changing each weight-$L$ edge to $3$ while filling the diagonals with $2$ achieves $\cdist(A)=|H(A)|/n^2=1/(8m)$ on $n=4m$ vertices. Setting $m=\lceil 1/(8\theta)\rceil$ completes the proof.
\end{proof}

\subsubsection*{An entropy inequality on cycles}

To bound the hitting number of shortest nonmetric cycles when $C>3$, we establish an entropy inequality for random variables indexed by a cycle (using natural logarithms). The following conditional independence property is a cycle separator form of the global Markov property for maximum entropy graphical models~\cite{DLS,Lauritzen} that also allows prescribed marginals with zero entries~\cite{Csiszar,WJ}.

\begin{lemma}\label{lem:arc-independence}
Let $X_0,\ldots,X_{r-1}$ be finite-valued random variables on a cycle with a feasible family of prescribed marginals on adjacent pairs. Any entropy-maximizing joint distribution with these marginals makes the interiors of complementary arcs conditionally independent given their two common endpoints.
\end{lemma}

\begin{proof}
An entropy maximizer $\mu$ exists by compactness. For separator endpoints $S$ and complementary arc interiors $A,B$, the conditional-product law $\nu(A,B,S)=\mu(S)\mu(A\mid S)\mu(B\mid S)$ preserves the marginals on $(A,S)$ and $(B,S)$ and hence all adjacent-pair marginals, while $H_\nu(A,B,S)-H_\mu(A,B,S)=I_\mu(A;B\mid S)\ge0$. Maximality forces $I_\mu(A;B\mid S)=0$.
\end{proof}

\begin{lemma}\label{lem:cyclic-entropy-inequality}
Let $r\ge4$ and suppose finite-valued random variables $(X_0,\ldots,X_{r-1})$ satisfy conditional arc independence on the cycle. With cyclic indices, the average pair entropies $E_d=\frac1r\sum_{i=0}^{r-1}H(X_i,X_{i+d})$ satisfy $E_d\ge E_1$ for all $2\le d\le r-2$.
\end{lemma}

\begin{proof}
Fix $i$ and $2\le d\le r-2$, and write $(A,B,C,D)=(X_i,X_{i+1},X_{i+d},X_{i+d+1})$. Conditional arc independence gives $A\perp C\mid(B,D)$ and $B\perp D\mid(A,C)$, so $H(A,B,D)+H(B,C,D)=H(A,B,C,D)+H(B,D)$ and $H(A,B,C)+H(A,C,D)=H(A,B,C,D)+H(A,C)$. Expanding the nonnegative sum $I(A;B\mid C,D)+I(B;C\mid A,D)+I(C;D\mid A,B)+I(D;A\mid B,C)\ge0$ therefore yields
$2H(A,C)+2H(B,D) \ge H(A,B)+H(B,C)+H(C,D)+H(D,A)$.
Averaging over $i$ gives $4E_d\ge 2E_1+E_{d-1}+E_{d+1}$. Since $E_{r-1}=E_1$, the differences $g_d=E_d-E_1$ satisfy $g_1=g_{r-1}=0$ and $4g_d\ge g_{d-1}+g_{d+1}$, which rules out a negative interior minimum.
\end{proof}

Unlike projection inequalities such as Shearer's inequality~\cite{CGFS}, \Cref{lem:cyclic-entropy-inequality} compares distinct averages of pair entropies and requires conditional arc independence: on a four-cycle with $X_0=X_2=U$ and $X_1=X_3=V$ for independent fair bits $U,V$, one has $E_1=2\log2>E_2=\log2$.

\begin{lemma}\label{lem:hitting-cycles}
Let $G$ be a positively weighted simple graph on $n$ vertices with $h=\binom n2-|E(G)|$ holes. If $G$ has no nonmetric cycle of length less than $r$, where $r\ge4$, then there is a set of at most $rh$ edges meeting every nonmetric $r$-cycle.
\end{lemma}

\begin{proof}
For each cyclic $r$-tuple $\omega=(v_0,\ldots,v_{r-1})$ of adjacent vertices in $G$, define the excess $s(\omega)=w(v_{r-1}v_0)-\sum_{i=0}^{r-2}w(v_iv_{i+1})$, and let $\Omega=\{\omega:s(\omega)>0\}$. Assume $\Omega\ne\varnothing$. Because $G$ has no nonmetric cycle of length less than $r$, every $\omega\in\Omega$ is a simple chordless cycle: a repeated vertex would yield a shorter closed subwalk and simple path of weight less than $w(v_{r-1}v_0)$, and a chord would split $\omega$ into two shorter cycles at least one of which is nonmetric. In particular, every two-step pair $(v_i,v_{i+2})$ along $\omega\in\Omega$ is a hole, so $h>0$.

Over the compact feasible polytope $\bigl\{z\in\mathbb R_{\ge0}^\Omega:\sum_{\omega:e\in E(\omega)}z_\omega\le1\text{ for all }e\in E(G)\bigr\}$, choose a fractional packing $z$ that first maximizes $\sum_{\omega\in\Omega}s(\omega)z_\omega>0$ and second minimizes the total mass $t=\sum_{\omega\in\Omega}z_\omega>0$. By averaging with cycle reversal $(v_0,\ldots,v_{r-1})\mapsto(v_{r-1},\ldots,v_0)$, we may assume $z$ is reversal-invariant. Let $F$ be the set of saturated edges; then $F$ meets every $\omega\in\Omega$ by score maximality, and $|F|\le\sum_{e\in E(G)}\sum_{\omega:e\in E(\omega)}z_\omega=rt$.

Normalize $\mu(\omega)=z_\omega/t$, and let $\nu$ be the entropy-maximizing distribution on $[n]^r$ matching the ordered adjacent-pair marginals of $\mu$. Then $\nu$ is supported on closed $r$-walks in $G$, satisfies $t\sum_{i=0}^{r-1}\Pr_\nu[\{X_i,X_{i+1}\}=e]\le1$ for every $e\in E(G)$, and has expected score $t\,\mathbb E_\nu s=\sum_{\omega\in\Omega}s(\omega)z_\omega$. Since positive-score walks lie in $\Omega$ and are simple, $z'_\omega=t\nu(\omega)$ ($\omega\in\Omega$) defines a feasible packing. If $\nu$ placed positive mass outside $\Omega$, restricting to $\Omega$ would either strictly increase the score (if $\nu(s<0)>0$) or preserve the optimal score while strictly reducing $t$ (if $\nu(s=0)>0$), contradicting the choice of $z$. Thus $\nu$ is supported on $\Omega$.

By reversal invariance, $p_i(u,v)=\Pr_\nu[(X_i,X_{i+1})=(u,v)]$ satisfies $p_i(u,v)=p_{r-2-i}(v,u)$. Because $\nu$ is supported on simple cycles, $u\ne v$, so $(X_i,X_{i+1})=(u,v)$ and $(X_{r-2-i},X_{r-1-i})=(v,u)$ are distinct ordered occurrences of the undirected edge $\{u,v\}$ even when the cyclic indices $i$ and $r-2-i$ coincide. Hence $2p_i(u,v)\le\sum_{j=0}^{r-1}\Pr_\nu[\{X_j,X_{j+1}\}=\{u,v\}]\le 1/t$, giving $H(X_i,X_{i+1})\ge\log(2t)$. On the other hand, because every $\omega\in\Omega$ is chordless, $(X_i,X_{i+2})$ takes values in the $2h$ ordered holes, so $H(X_i,X_{i+2})\le\log(2h)$. Lemmas~\ref{lem:arc-independence} and~\ref{lem:cyclic-entropy-inequality} therefore yield
$\log(2t) \le E_1 \le E_2 \le \log(2h)$,
so $t\le h$ and $|F|\le rh$.
\end{proof}

\begin{theorem}[Completion stability for metrics with bounded aspect ratio]\label{thm:completion-stability-C}
For fixed $C>1$, define $M_C=\prod_{r=4}^{\lceil C\rceil}(r+1)$, where the empty product is $1$. Every locally consistent partial matrix in $[1,C]$ with $h$ holes becomes metrically completable after deleting at most $(M_C-1)h$ specified edges. Consequently,
$$
\compmod(\theta)\le\min\left\{1,2(M_C-1)\theta\right\},
$$
and for every fixed $C>3$, we have $\compmod(\theta)=\Theta_C(\theta)$ as $\theta\downarrow0$.
\end{theorem}

\begin{proof}
Local consistency excludes nonmetric triangles, and any nonmetric $r$-cycle must satisfy $r-1<C$, so $r\le\lceil C\rceil$. For $r=4,\ldots,\lceil C\rceil$ in turn, delete a hitting set of specified edges for all nonmetric $r$-cycles. Because edge deletion creates no new cycles, each stage $r$ has no nonmetric cycle of length less than $r$; if $h_r$ is the current hole count, Lemma~\ref{lem:hitting-cycles} deletes at most $rh_r$ edges, giving $h_{r+1}\le(r+1)h_r$. After stage $\lceil C\rceil$, the hole count is at most $M_Ch$, at most $(M_C-1)h$ specified edges have been deleted, and no nonmetric cycle remains, so Lemma~\ref{lem:metric-completion-criterion} provides an exact completion. Normalizing the edit count yields the upper bound, while Proposition~\ref{prop:exact-small-c} gives $\compmod\equiv0$ for $C\le3$ and the matching linear lower bound for $C>3$.
\end{proof}

\subsubsection*{Uniform polynomial-time completion and metric repair}

We next show that the hitting sets of Lemma~\ref{lem:hitting-cycles} and the resulting completions can be computed in polynomial time via linear programming duality.

\begin{lemma}\label{lem:cycle-hitting-algorithm}
Under the hypotheses of Lemma~\ref{lem:hitting-cycles}, if the edge weights are rational, a hitting set of size at most $rh$ can be computed in time polynomial in $n$ and the input bit length, with polynomial degree independent of $r$.
\end{lemma}

\begin{proof}
If $r>n$, return $\varnothing$. Otherwise, consider the dual of the maximum-score packing linear program: minimize $\sum_{e\in E(G)}y_e$ subject to $y_e\ge0$ and $\sum_{e\in E(\omega)}y_e\ge s(\omega)$ for all $\omega\in\Omega$. For a candidate $y\ge0$ and each edge $e=uv$, dynamic programming computes the minimum of $\sum_{f\in P}(w_f+y_f)$ over $u$-to-$v$ walks $P$ of at most $r-1$ edges. A constraint with dominant edge $e$ is violated if and only if this minimum is strictly less than $w_e-y_e$: any violating walk has original weight less than $w_e$, and shortcutting repeated vertices preserves the strict inequality (since $w_f+y_f>0$) and yields a nonmetric simple cycle of length at most $r$, which must have length $r$ by hypothesis.

Since $y_e=w_e$ is feasible, we may restrict the dual to a rational box of polynomial bit complexity and compute an exact optimum $y^*$ in polynomial time via the ellipsoid method~\cite{GLS}. If $\Omega=\varnothing$, then $y^*=0$ and $\varnothing$ suffices. Otherwise, let $z$ be the maximum-score, minimum-mass primal optimum from the proof of Lemma~\ref{lem:hitting-cycles}. Complementary slackness implies $\sum_{\omega:e\in E(\omega)}z_\omega=1$ whenever $y^*_e>0$, so $|\supp y^*|\le r\sum_{\omega\in\Omega}z_\omega\le rh$, and every $\omega\in\Omega$ meets $\supp y^*$ because $s(\omega)>0$.
\end{proof}

\begin{corollary}\label{cor:algorithmic-completion-repair}
For every real $C>1$, a locally consistent partial matrix with rational specified entries in $[1,C]$ and $h$ holes admits a rational metric completion in $[1,C]$ after changing at most $(M_C-1)h$ specified entries, computable in time polynomial in $n$ and the input bit length with degree independent of $C$ (and without knowledge of $C$).
\end{corollary}

\begin{proof}
Set $R=\max\bigl(\{1\}\cup\{\text{specified entries}\}\bigr)\le C$. If $R=1$, output the all-ones off-diagonal matrix. Otherwise apply Lemma~\ref{lem:cycle-hitting-algorithm} for $4\le r\le\min\{n,\lceil R\rceil\}$ and complete the remaining graph via $B_{uv}=\min\{R,d_G(u,v)\}$ for $u\ne v$.
\end{proof}

For $n\ge2$ and $N=\binom n2$, write $\dist_{\mathrm H}(D,\mathcal M_n) = \frac1N\min_{d\in\mathcal M_n}|\{e:D_e\ne d_e\}| = \frac n{n-1}\dist_n(D,\mathcal M_n)$ for the edge-normalized Hamming distance. Clipping an arbitrary pseudometric to $[1,C]$ preserves both the triangle inequality and agreement with entries in $[1,C]$, yielding the following transfer principle.

\begin{lemma}\label{lem:clipping-equivalence}
For every $D\in[1,C]^{\binom{[n]}2}$, we have $\dist_n(D,\mathcal M_n[1,C]) = \dist_n(D,\mathcal M_n) = \dist_n(D,\overline{\mathcal M}_n)$.
\end{lemma}

\begin{proof}
For any $d\in\overline{\mathcal M}_n$, setting $\widehat d_{uv}=\min\{C,\max\{1,d_{uv}\}\}$ for $u\ne v$ and $\widehat d_{uu}=0$ preserves the triangle inequality (since $\max\{1,d_{uw}\}+\max\{1,d_{wv}\}\ge\max\{2,d_{uw}+d_{wv}\}\ge\max\{1,d_{uv}\}$ and capping at $C$ preserves the triangle inequality as in Lemma~\ref{lem:metric-completion-criterion}) without altering agreement with $D\in[1,C]^{\binom{[n]}2}$.
\end{proof}

Combining \Cref{thm:completion-stability-C} with \Cref{prop:repair-ineq} and \Cref{lem:clipping-equivalence} bounds the metric repair distance on $[1,C]$ within a factor of $3M_C$ of the maximum size $\nu(D)$ of an edge-disjoint packing of violating triangles.

\begin{proposition}[Triangle packing bound for metric repair]\label{prop:metric-repair-approx}
For $n\ge3$ and $D\in[1,C]^{\binom{[n]}2}$,
$$
\frac{2\nu(D)}{n^2}
\le
\dist_n(D,\mathcal M_n)
=
\dist_n(D,\mathcal M_n[1,C])
\le
\frac{6M_C\nu(D)}{n^2}.
$$
\end{proposition}

\begin{proof}
The lower bound is Lemma~\ref{lem:packing-viol-edits}. For the upper bound, deleting all $3p\le 3\nu(D)$ edges of a maximal edge-disjoint family of violating triangles leaves a locally consistent partial matrix, which Theorem~\ref{thm:completion-stability-C} completes after at most $3(M_C-1)p$ further deletions; Lemma~\ref{lem:clipping-equivalence} equates the repair targets.
\end{proof}

\begin{remark}[Support selection and bounded cycle approximation]\label{prop:cycle-approx}
When $C\le3$, Proposition~\ref{prop:exact-small-c} implies that the minimum number of metric repair edits equals the minimum size of an edge set hitting all violating triangles. Selecting an optimal hitting set is nevertheless NP-hard and APX-hard even for weights in $\{1,2,3\}$~\cite{FRV} (see also~\cite{BLLSSSX} for parameterized kernels). For any fixed $C>2$, repeatedly deleting the edges of a shortest nonmetric cycle (found via shortest paths and having length at most $\lceil C\rceil$) and completing via Corollary~\ref{cor:algorithmic-completion-repair} recovers the polynomial-time $\lceil C\rceil$-approximation specialization of~\cite{FRV,FGRSV}, complementing the range-independent $O(\log n)$-approximation of Cohen-Addad, Fan, Lee, and de Mesmay~\cite{CFLM}.
\end{remark}

\subsection{Twin rigidity, completion obstructions, and logical independence}\label{sec:completion-obstructions}

Whereas our pseudometric repairs in Lemmas~\ref{lem:tree-completion-stability} and~\ref{lem:completion-stability-euclidean-pseudo} may identify distinct vertices, injective realization in trees or $\mathbb R^d$ forbids zero off-diagonal distances. We now reformulate the deterministic collinear twin pair construction of Parnas and Ron~\cite[Section~7]{PR} as a locally consistent partial matrix with $m$ holes to show that injectivity creates a completion obstruction for $\mathcal T^+$ and $\mathcal E^{(d),+}$.

\begin{lemma}\label{lem:twin-rigidity}
Let $m\ge3$ and $y_1<\cdots<y_m$. On $V=[m]\times[2]$, write $i^a=(i,a)$, leave the $m$ pairs $\{i^1,i^2\}$ within each class unspecified as holes, and set $A_{i^a,j^b}=|y_i-y_j|$ for $i\ne j$. Every complete matrix $B\in\mathcal T_{2m}^+\cup\mathcal E_{2m}^{(d),+}$ differs from $A$ on at least $m(m-1)/3$ specified edges, and hence satisfies $\cdist(A)\ge\frac{m-1}{6m}$.
\end{lemma}

\begin{proof}
For $1\le j<i<k\le m$ and $a,b\in\{1,2\}$, consider the four-vertex witness $W=\{j^a,k^b,i^1,i^2\}$. If $B$ agrees with $A$ on all five specified edges of $W$, then $B_{j^a,i^1}=B_{j^a,i^2}=y_i-y_j$, $B_{i^1,k^b}=B_{i^2,k^b}=y_k-y_i$, and $B_{j^a,k^b}=y_k-y_j$. In $\mathcal T_{2m}^+$, the three opposite-edge sums on $W$ are $y_k-y_j+B_{i^1,i^2}$ and twice $y_k-y_j$, so the four-point condition forces $B_{i^1,i^2}\le0$, a contradiction. In $\mathcal E_{2m}^{(d),+}$, equality in the triangle inequality on $\{j^a,i^1,k^b\}$ and $\{j^a,i^2,k^b\}$ places both $\phi(i^1)$ and $\phi(i^2)$ at the unique point on the segment $[\phi(j^a),\phi(k^b)]$ at distance $y_i-y_j$ from $\phi(j^a)$, forcing $B_{i^1,i^2}=0$, again a contradiction.

Consequently, $B$ must differ from $A$ on at least one specified edge in each of the $4\binom m3$ witnesses. A specified edge between classes $p<q$ belongs to $2(p-1)+2(m-q)+(q-p-1)\le2(m-2)$ witnesses, so double counting yields at least $4\binom m3/(2(m-2))=m(m-1)/3$ specified discrepancies, or $\cdist(A)\ge(m-1)/(6m)$ after normalizing by $2/(2m)^2$.
\end{proof}

\begin{theorem}[Injective tree and Euclidean completion obstructions]\label{thm:completion-obstructions}
The strict predicates $\Phi_{\mathrm{tree}}^+$ and $\Phi_{\mathrm{Euc}}^{(d),+}$ admit completion obstructions and are not removable. Consequently, neither $\mathcal T^+$ nor $\mathcal E^{(d),+}$ has a constant-query one-sided tester.\footnotemark
\end{theorem}
\footnotetext{Although Bao, Kannan, and Waingarten~\cite{BKW} claimed a constant-query upper bound on testing of strict tree metrics [Lemma 5.7, pg. 25-26], their upper bound proofs implicitly rely on non-injectivity.}

\begin{proof}
Apply Lemma~\ref{lem:twin-rigidity} with $y_i=i$ and $m\ge\max\{6,d+3\}$. Every fully specified subset contains at most one vertex per twin class and embeds injectively on a line, while the hole density is $m/(2m)^2=1/(4m)\to0$ and $\cdist(A)\ge(m-1)/(6m)\ge1/8$. Corollaries~\ref{cor:obstructions} and~\ref{cor:test-structure} complete the proof.
\end{proof}

\begin{remark}[Geodesic rigidity, injectivity, and two-sided testing]\label{rem:strict-ultrametrics}
Filling the $m$ twin holes in Lemma~\ref{lem:twin-rigidity} with zeros gives an exact non-injective line realization, showing that the completion obstruction arises from injectivity rather than from graph non-chordality (cf.~\cite{BJ}). Specifically, in a tree or in $\mathbb R^d$, any intermediate point $x$ satisfying $d(u,x)=a$ and $d(x,v)=d(u,v)-a$ for $0<a<d(u,v)$ is uniquely determined, forcing twin vertices to coincide. The same double-counting argument (with $b+1$ clones per class) applies in any metric space where this two-distance system has at most $b$ solutions along a geodesic segment; by contrast, in $(\mathbb R^2,\|\cdot\|_\infty)$, where infinitely many points satisfy the system simultaneously, twin vertices can be separated to positive distance without altering cross-class entries.

Ultrametrics avoid geodesic rigidity altogether: if $0<a<d(u,v)$, then $\max\{a,d(u,v)-a\}<d(u,v)$, so no point satisfies the two-distance system above, and twin holes over a positive ultrametric can be filled by any sufficiently small positive value. Accordingly, strict ultrametrics satisfy both packing amplification and completion stability on $\mathbb R_{\ge0}$ and $(0,\infty)$ (Corollaries~\ref{cor:strict-packing} and~\ref{cor:strict-ult-completion}); in fact, for any nonnegative input $D$ with $z$ off-diagonal zero entries, replacing zeros in an optimal repair $U\in\mathcal U_n$ by a small $\delta>0$ below all positive entries of $D$ and $U$ shows that $\dist_n(D,\mathcal U_n^+)\le\dist_n(D,\mathcal U_n)+2z/n^2$. Thus injectivity alone does not create a completion obstruction. Finally, for $\mathcal T^+$ and $\mathcal E^{(d),+}$, randomizing twin pairs on parallel lines or comb trees yields $\Omega(\sqrt n)$ two-sided testing lower bounds~\cite[Theorem~9]{PR} (see also~\cite{KS,Onak}).
\end{remark}

Finally, combining our results from Sections~\ref{sec:packing} and~\ref{sec:completion} demonstrates that neither structural condition in Theorem~\ref{thm:structure} implies the other.

\begin{corollary}[Logical independence of the structural conditions]\label{cor:independence}
Packing amplification does not imply completion stability, and completion stability does not imply packing amplification.
\end{corollary}

\begin{proof}
For $\Lambda=[1,C]$ with $C>2$, the metric predicate $\Phi_{\mathrm{met}}$ is completion stable by Theorem~\ref{thm:completion-stability-C} (with $\compmod\equiv0$ for $2<C\le3$ by Proposition~\ref{prop:exact-small-c}), but fails packing amplification by Theorem~\ref{thm:metric-packing-obstruction}. Conversely, on $\Lambda=\mathbb R_{\ge0}$ or $\Lambda=(0,\infty)$, the strict predicates $\Phi_{\mathrm{tree}}^+$ and $\Phi_{\mathrm{Euc}}^{(d),+}$ satisfy packing amplification by Corollary~\ref{cor:strict-packing}, but fail completion stability by Theorem~\ref{thm:completion-obstructions}.
\end{proof}

\section{Quantitative testing bounds}\label{sec:quantitative-testing}

We establish quantitative query complexity bounds for testing our geometric and metric families. We first combine our packing and completion bounds from Sections~\ref{sec:packing} and~\ref{sec:completion} to obtain explicit constant-query one-sided testers for all four removable geometric families: ultrametrics, strict ultrametrics, tree pseudometrics, and $d$-dimensional Euclidean pseudometrics (Corollary~\ref{cor:geometric-testing}). Turning to general metrics, for which Bao, Kannan, and Waingarten~\cite{BKW} proved a nonadaptive one-sided upper bound of $O(n^{2/3}/\epsilon^{4/3})$ and a nonadaptive one-sided lower bound of $\Omega(n^{2/3-o(1)})$ at vanishing proximity $\epsilon = n^{-o(1)}$, we introduce a signed noise perturbation scheme on robust metric faces (\Cref{lem:metric-noise,lem:metric-two-edge,lem:metric-hybrid}) and pair it with density sensitive submatrix sampling and asymmetric faces to determine the sharp nonadaptive one-sided and two-sided query complexity $\Theta_C(F(n,\epsilon))$ on $[1,C]$ for $C>2$ across all four proximity regimes of $(n,\epsilon)$ (\Cref{thm:sharp-nonadaptive-metric}). Next, we introduce a contextual blow-up of our robust faces that prevents adaptive coordinate recovery, establishing an adaptive two-sided lower bound of $\Omega_C(\sqrt{n/\log n})$ queries (Theorem~\ref{thm:metric-adaptive}). Finally, we apply our clipping equivalence (Lemma~\ref{lem:clipping-equivalence}) to transfer both the nonadaptive two-sided lower bound $\Omega(F(n,\epsilon))$ and the adaptive lower bound $\Omega(\sqrt{n/\log n})$ to unrestricted metrics and pseudometrics (Theorem~\ref{thm:unrestricted-metric-lower}).

\subsection{Quantitative testing of removable geometric families}

Combining Proposition~\ref{prop:repair-ineq} and Theorem~\ref{thm:testing} with the bounds of Sections~\ref{sec:packing} and~\ref{sec:completion} gives explicit query bounds for all four removable geometric families.

\begin{corollary}[Quantitative geometric testers]\label{cor:geometric-testing}
For every $0<\epsilon\le1$, the removal moduli of $\mathcal U$, $\mathcal T$, and $\mathcal E^{(d)}$ satisfy
$$
\rho_{\mathrm{ult}}(\epsilon)\ge\frac{\epsilon^2}{9600},
\qquad
\rho_{\mathrm{tree}}(\epsilon)
\ge
\min\left\{\frac1{1820},
\frac3{1600}\left(\frac{\epsilon}{168}\right)^2\right\},
\qquad
\rho_{\mathrm{Euc}}^{(d)}(\epsilon)
\ge
\min\left\{\delta_d,
\left(\frac{\epsilon}{C_d}\right)^{d+2}\right\},
$$
where $C_d$ and $\delta_d$ are as in Lemma~\ref{lem:packing-euclidean-pseudo}; for strict ultrametrics $\mathcal U^+$, we have $\rho_{\mathrm{ult}^+}(\epsilon)\ge\epsilon^2/9600$ on $(0,\infty)$ and $\rho_{\mathrm{ult}^+}(\epsilon)\ge\epsilon^2/86400$ on $\mathbb R_{\ge0}$. Consequently, $\mathcal U$, $\mathcal U^+$, and $\mathcal T$ admit one-sided $O(\epsilon^{-2})$-query $\epsilon$-testers, and $\mathcal E^{(d)}$ admits a one-sided $O_d(\epsilon^{-(d+2)})$-query $\epsilon$-tester.
\end{corollary}

\begin{proof}
By Proposition~\ref{prop:repair-ineq}, Lemma~\ref{lem:ult-completion} gives $\dist_n(D,\mathcal U_n)\le24\nu(D)/n^2$, Corollary~\ref{cor:strict-ult-completion} gives $\dist_n(D,\mathcal U_n^+)\le36\nu_{\mathrm{ult}^+}(D)/n^2$ on $\mathbb R_{\ge0}$, and Lemma~\ref{lem:tree-completion-stability} gives $\dist_n(D,\mathcal T_n)\le168\nu(D)/n^2$. Substituting the resulting packing lower bounds $\alpha=\epsilon/24$, $\epsilon/36$, and $\epsilon/168$ into Lemmas~\ref{lem:ult-packing} and~\ref{lem:tree-packing} and Proposition~\ref{prop:pseudometric-to-metric-packing} (with $\packmod_{\mathrm{ult}^+}(\alpha)\ge 3\alpha^2/200$) yields the stated moduli for $\mathcal U$, $\mathcal U^+$, and $\mathcal T$, while Corollary~\ref{cor:euc-removal} gives the modulus for $\mathcal E^{(d)}$. Theorem~\ref{thm:testing} converts each removal modulus into the claimed one-sided query bound.
\end{proof}

\begin{remark}[Comparison with prior geometric testing bounds]
For ultrametrics and tree pseudometrics, Parnas and Ron~\cite{PR} gave nonadaptive $O(\epsilon^{-6})$-query testers, which Bao, Kannan, and Waingarten~\cite{BKW} improved to $\widetilde O(\epsilon^{-2})$ queries on $\widetilde O(\epsilon^{-1})$ sampled vertices alongside an $\Omega(\epsilon^{-1})$ vertex sample lower bound and an $\Omega(\epsilon^{-4/3})$ two-sided query lower bound. Corollary~\ref{cor:geometric-testing} removes the logarithmic factors from the query bound of~\cite{BKW} via uniform $k$-set sampling (and applying our linear completion stability bounds to an induced submatrix on $O(\epsilon^{-1}\log(1/\epsilon))$ vertices likewise recovers their single-submatrix sample bound). For $d$-dimensional Euclidean pseudometrics, our $O_d(\epsilon^{-(d+2)})$ bound reflects the $(d+3)$-point proximity-oblivious~\cite{GR} removal modulus; when larger submatrices are queried, better $\epsilon$-dependence is achieved by the $\widetilde O(d^2/\epsilon^2)$-query tester of~\cite{PR} and the $O(d^2/\epsilon^2)$-query tester of Krauthgamer and Sasson~\cite{KS} (see also Onak~\cite{Onak}).
\end{remark}

\subsection{Signed noise on robust metric faces}

To convert our edge-disjoint tight-triangle constructions into two-sided testing lower bounds, we introduce a real-valued additive counterpart to the $\mathbb F_2$-parity lifting of Alon, Kaufman, Krivelevich, and Ron~\cite[Section~6.1]{AKKR} for triangle-freeness. Working directly on weighted matrices without vertex blow-up, we perturb the three edges of each designated tight triangle by correlated uniform shifts $(U,V,U+V\mp 1)\eta$ over $\{-M,\ldots,M\}\eta$. Because any two edges of a selected triangle have nearly identical joint distributions under the two signs (an additive analogue of the Independence Observation of~\cite{AKKR}), a stopped transcript hybrid argument reduces two-sided distinguishing to the exposure event $F_{\mathcal A}$ of querying all three edges of some selected triangle.

\begin{lemma}\label{lem:metric-noise}
Suppose a rational metric $A\in[1,C_0]^{\binom{[n]}2}$ has an edge-disjoint family $\mathcal H$ of $h$ designated tight triangles, every other triangle inequality has slack at least $n^{-K}$, every selected edge has endpoint margin at least $n^{-K}$, and all entries have $O_C(\log n)$-bit rational representations for a fixed $K$. Fix an integer $J>K+2$ and set $\delta=n^{-J}/1000$, $M=n^{10}$, and $\eta=\delta/M$. There exist finite-support distributions $\mu_-$ and $\mu_+$, obtained by independent triangle perturbations followed by uniform vertex relabeling, such that:
\begin{enumerate}[label=(\arabic*)]
\item all entries lie in $[1,C_0]$ and have $O_C(\log n)$-bit rational representations (with constants allowed to depend on $K,J$);
\item every instance in $\supp\mu_-$ is metric;
\item before relabeling, every instance in $\supp\mu_+$ violates exactly the designated tight inequalities of $\mathcal H$;
\item every instance $D\in\supp\mu_+$ satisfies $\dist_n(D,\mathcal M_n)=2h/n^2$ and $\dist_{\mathrm H}(D,\mathcal M_n)=h/N$.
\end{enumerate}
\end{lemma}

\begin{proof}
For each triangle $H\in\mathcal H$, independently draw $U_H,V_H$ uniformly from $\{-M,-M+1,\ldots,M\}\eta$, add $U_H$ and $V_H$ to the two short edges of $H$, and add $U_H+V_H\mp\eta$ to the designated long edge under $\mu_\mp$. Each entry changes by at most $3\delta$ and each formerly strict slack changes by at most $9\delta<n^{-K}$, while the designated slack on each $H\in\mathcal H$ becomes $\pm\eta$. Thus $\mu_-$ is supported on metrics and $\mu_+$ violates precisely the $h$ edge-disjoint triangles in $\mathcal H$, requiring exactly $h$ edge edits (achieved by restoring the $h$ long edges under $\mu_-$).
\end{proof}

\begin{lemma}\label{lem:metric-two-edge}
For a single triangle in Lemma~\ref{lem:metric-noise}, any adaptive procedure observing at most two of its edges (even with sign-independent side information) has transcript total variation distance at most $6/(2M+1)$ between $\mu_-$ and $\mu_+$. For a fixed projection onto at most two edges, the total variation distance is at most $2/(2M+1)$.
\end{lemma}

\begin{proof}
After subtracting the baseline entries and dividing by $\eta$, the three triangle coordinates are $(U,V,U+V-1)$ under $\mu_-$ and $(U,V,U+V+1)$ under $\mu_+$, where $U,V\in\{-M,\ldots,M\}$ are independent and uniform. The marginal on $(U,V)$ is identical under both signs, while conditional on $U$ (or on $V$) the third coordinate is uniform on an interval shifted by $2$, giving total variation $2/(2M+1)$. Conditioning on the side information and internal randomness fixes the first queried coordinate, and a union bound over the three possible two-coordinate pairs bounds the adaptive transcript variation by $6/(2M+1)$.
\end{proof}

\begin{lemma}\label{lem:metric-hybrid}
Let $\mathcal A$ query the distributions of Lemma~\ref{lem:metric-noise}, and let $F_{\mathcal A}$ be the event that $\mathcal A$ queries all three edges of some selected triangle in $\mathcal H$. Then
$$
\TV\bigl(
\Pi_{\mathcal A}(\mu_-),
\Pi_{\mathcal A}(\mu_+)
\bigr)
\le
\Pr_{\mu_-}(F_{\mathcal A})
+
\Pr_{\mu_+}(F_{\mathcal A})
+
\frac{6h}{2M+1},
$$
and the same bound holds when $(A,\mathcal H)$ is drawn from a sign-independent distribution with $|\mathcal H|=h$ almost surely. If $\mathcal A$ is nonadaptive, then $\TV\bigl(\Pi_{\mathcal A}(\mu_-),\Pi_{\mathcal A}(\mu_+)\bigr) \le \Pr(F_{\mathcal A})+\frac{2h}{2M+1}$, where $F_{\mathcal A}$ depends only on the query set, base template, and relabeling.
\end{lemma}

\begin{proof}
Condition on the base template and relabeling, and modify $\mathcal A$ to halt immediately before any query that would reveal the third edge of a triangle in $\mathcal H$. Replacing the $h$ independent triangle laws one by one changes the stopped transcript distribution by at most $6/(2M+1)$ per triangle by Lemma~\ref{lem:metric-two-edge}, and the unstopped transcript agrees with the stopped transcript outside $F_{\mathcal A}$ under each sign. For nonadaptive $\mathcal A$, conditioning on the random seed fixes the queried edges; outside $F_{\mathcal A}$, each triangle is observed on a fixed set of at most two edges, so the fixed-projection bound of Lemma~\ref{lem:metric-two-edge} gives $2h/(2M+1)$.
\end{proof}

\subsection{Density-sensitive sampling and asymmetric faces}

To match our nonadaptive lower bounds across all proximity regimes, we pair a second-moment submatrix sampling bound on the upper-bound side with asymmetric and single-center robust face constructions on the lower-bound side.

\begin{lemma}\label{lem:metric-sample-sharp}
Let $\mathcal S$ be an edge-disjoint family of $m\ge\gamma n^2$ triangles on $[n]$. For a uniformly random $r$-set $U\subseteq[n]$ with $6\le r<n$, we have
$\Pr[\exists T\in\mathcal S:T\subseteq U] \ge \bigl(\frac{8n}{\gamma r^3} + 1 + \frac{12}{r}\min\{\frac1{2\gamma},n\}\bigr)^{-1}$.
In particular, this probability is at least $2/5$ whenever $6\le r<n$, $r\ge2(n/\gamma)^{1/3}$, and $r\ge12/\gamma$.
\end{lemma}

\begin{proof}
Let $X=|\{T\in\mathcal S:T\subseteq U\}|$ and $p_j=(r)_j/(n)_j$, so $\mu=\mathbb EX=mp_3\ge\gamma n^2(r-2)^3/n^3\ge\gamma r^3/(8n)$. Each vertex belongs to at most $\Delta\le\min\{(n-1)/2,m\}$ triangles of $\mathcal S$, so there are at most $3m\Delta$ ordered pairs of intersecting triangles. Since vertex-disjoint triangles have joint inclusion probability $p_6\le p_3^2$, we have $\mathbb EX^2\le\mu+\mu^2+3m\Delta p_5$. Using $p_5/p_3^2\le 4n/r$ for $n>r\ge6$ gives $\mathbb EX^2/\mu^2 \le \frac{8n}{\gamma r^3}+1+\frac{12\Delta n}{rm} \le \frac{8n}{\gamma r^3}+1+\frac{12}{r}\min\{\frac1{2\gamma},n\}$, and the Paley--Zygmund inequality $\Pr[X>0]\ge\mu^2/\mathbb EX^2$ gives the result.
\end{proof}

\begin{lemma}\label{lem:asymmetric-face}
Fix $C>2$ and let $C_0$ be as in Lemma~\ref{lem:metric-face}. For sufficiently large $n$, put $m=\lfloor n/4\rfloor$. Given either (i) $1\le a\le m$ and $h=am$, or (ii) $a=1$ and any $1\le h\le\lfloor(n-1)/2\rfloor$, there is a rational metric $A\in[1,C_0]^{\binom{[n]}2}$ with an edge-disjoint family $\mathcal H$ of $h$ tight triangles such that:
\begin{enumerate}[label=(\arabic*)]
\item every $T\in\mathcal H$ contains one of $a$ designated centers, and the triangles of $\mathcal H$ through each center form a matching of size $h/a$ on the remaining vertices;
\item all other triangle inequalities and endpoint margins of selected edges are at least $n^{-3}$, and all entries are $O_C(\log n)$-bit rationals.
\end{enumerate}
\end{lemma}

\begin{proof}
For $h=am$ with $1\le a\le m$, take the construction of Lemma~\ref{lem:metric-face} on $4m-1$ vertices, retain $a$ vertices of $Y$ together with all of $X$ and $Z$, and pad to $n$ vertices with distance $3/2$; for each retained $y_j$, the pairs $\{x_{k-j},z_k\}$ ($k\in[m]$) form a matching of size $m$. For a single center with $1\le h\le\lfloor(n-1)/2\rfloor$, set $b=1+4\rho$ and $C_0=2+10\rho$ as in Lemma~\ref{lem:metric-face}, assign weight $b$ to the $2h$ edges from the center to the leaves, weight $2b$ to the $h$ edges between matched leaves, and weight $3/2$ to all remaining edges; the $h$ matched triangles satisfy $b+b=2b$ and all other slacks and endpoint margins are bounded below by positive constants depending only on $C$.
\end{proof}

\subsection{The full nonadaptive complexity on a bounded range}\label{sec:full-nonadaptive}

Write $Q_{\mathrm{na}}^{(1)}(n,\epsilon;C)$ and $Q_{\mathrm{na}}^{(2)}(n,\epsilon;C)$ for the nonadaptive one-sided and two-sided query complexities of metric testing on $[1,C]$. For $n\ge3$ and $0<\epsilon\le1$, define the rate function
$$
F(n,\epsilon)
=
\begin{cases}
n^2,&0<\epsilon\le n^{-2},\\[1mm]
n/\sqrt\epsilon,&n^{-2}\le\epsilon\le n^{-1},\\[1mm]
\sqrt n/\epsilon,&n^{-1}\le\epsilon\le n^{-1/2},\\[1mm]
(n/\epsilon)^{2/3},&n^{-1/2}\le\epsilon\le1.
\end{cases}
$$

\begin{theorem}[Sharp nonadaptive complexity on a known bounded range]\label{thm:sharp-nonadaptive-metric}
For every fixed $C>1$, we have $Q_{\mathrm{na}}^{(1)}(n,\epsilon;C)=O_C(F(n,\epsilon))$ for all $n\ge3$ and $0<\epsilon\le1$ (and $Q_{\mathrm{na}}^{(1)}(n,\epsilon;C)=0$ when $C\le2$). Conversely, for $\epsilon_0=1/80$ and every fixed $C>2$, there exist constants $A_C>0$, $n_0(C)$, and an absolute constant $c>0$ such that
$$
cF(n,\epsilon)
\le
Q_{\mathrm{na}}^{(2)}(n,\epsilon;C)
\le
Q_{\mathrm{na}}^{(1)}(n,\epsilon;C)
\le
A_CF(n,\epsilon)
$$
for all $n\ge n_0(C)$ and $0<\epsilon\le\epsilon_0$. Moreover, the hard instances have $O_C(\log n)$-bit rational entries, and the bounds hold whether repairs are restricted to $[1,C]$ or allowed arbitrary positive metric values.
\end{theorem}

\begin{proof}
When $C\le2$, every matrix in $[1,C]$ is metric. For $C>2$, set $a_C=6M_C$ and $\gamma=\epsilon/a_C$. By Proposition~\ref{prop:metric-repair-approx}, every $\epsilon$-far input has $\nu(D)>\gamma n^2$. By Lemma~\ref{lem:metric-sample-sharp}, for $6\le r<n$, repeating a uniform $r$-submatrix check $O\bigl(1+\frac{n}{\gamma r^3}+\frac1r\min\{\gamma^{-1},n\}\bigr)$ times yields a nonadaptive one-sided tester using $O\bigl(\frac{n}{\gamma r}+r^2+r\min\{\gamma^{-1},n\}\bigr)$ queries. For $n\ge36$, choosing $r$ up to rounding and truncation at $n$ (using full inspection whenever $n<36$, rounding reaches $n$, or $\gamma\le n^{-2}$) with $r\asymp (n/\gamma)^{1/3}$ for $\gamma\ge n^{-1/2}$, $r\asymp\sqrt n$ for $n^{-1}\le\gamma\le n^{-1/2}$, and $r\asymp\gamma^{-1/2}$ for $n^{-2}\le\gamma\le n^{-1}$ gives query complexity $O(F(n,\gamma))\le O(a_CF(n,\epsilon))$.

For the lower bound, we apply Lemmas~\ref{lem:metric-noise} and~\ref{lem:metric-hybrid} with $K=3$, $J=6$, and $M=n^{10}$ to three face families, each having $h\ge\epsilon n^2$ selected triangles so that every instance in $\supp\mu_+$ is at distance $2h/n^2\ge2\epsilon>\epsilon$ from $\mathcal M_n$ (and from $\mathcal M_n[1,C]$ by Lemma~\ref{lem:clipping-equivalence}). Fix the random seed of a nonadaptive $q$-query algorithm $\mathcal A$, let $G$ be its query graph of at most $q$ edges, and write $d_v$ and $t_v\le d_v^2/2$ for the degree and triangle count at vertex $v$. By the classical spectral bound~\cite{Alon} (used also in~\cite[Section~4]{BKW}), the total number of triangles $t(G)$ in $G$ satisfies $6t(G)=\operatorname{tr}(A_G^3)\le\bigl(\sum_i\lambda_i(A_G)^2\bigr)^{3/2}\le(2q)^{3/2}$, while $\sum_v d_v\le 2q$.

\emph{Case 1: $n^{-1/2}\le\epsilon\le\epsilon_0$.} With $\beta=1/20$, apply Lemma~\ref{lem:metric-face} on $s=\lceil n\sqrt{\epsilon/\beta}\rceil\le n/2+1$ vertices and pad to $n$ vertices with distance $3/2$. Then $\epsilon n^2\le\beta s^2\le h\le s^2/6\le \frac{2}{3\beta}\epsilon n^2$. Under uniform relabeling, each triangle of $G$ hits a selected triangle with probability $h/\binom n3$, so a union bound gives $\Pr(F_{\mathcal A})\le t(G)h/\binom n3 = O(\epsilon q^{3/2}/n) = O\bigl((q/F(n,\epsilon))^{3/2}\bigr)$.

\emph{Case 2: $1/(4n)\le\epsilon\le n^{-1/2}$.} Apply Lemma~\ref{lem:asymmetric-face} with $m=\lfloor n/4\rfloor$ and $a=\lceil 5\epsilon n\rceil\le 9\epsilon n$, so $h=am\ge\epsilon n^2$. Conditional on a designated center mapping to $v$, its $m$ selected leaf pairs form a uniform matching on $[n]\setminus\{v\}$, so the probability that $G$ contains a selected triangle through $v$ is at most $\min\{1,mt_v/\binom{n-1}2\}\le\min\{1,d_v^2/n\}\le d_v/\sqrt n$. Summing over $v$ and the $a$ centers gives $\Pr(F_{\mathcal A})\le \frac{a}{n}\sum_v \frac{d_v}{\sqrt n}\le 18\epsilon q/\sqrt n \le 18q/F(n,\epsilon)$ across both $n^{-1}\le\epsilon\le n^{-1/2}$ and $1/(4n)\le\epsilon\le n^{-1}$.

\emph{Case 3: $0<\epsilon<1/(4n)$.} Apply the single-center case of Lemma~\ref{lem:asymmetric-face} with $h=\lceil\epsilon n^2\rceil\le 2\max\{1,\epsilon n^2\}$. Conditioning on the center mapping to $v$ and bounding the matching intersection probability by $\min\{1,ht_v/\binom{n-1}2\}\le\min\{1,4hd_v^2/n^2\}\le 2\sqrt h\,d_v/n$ gives $\Pr(F_{\mathcal A})\le \frac1n\sum_v \frac{2\sqrt h\,d_v}{n}\le 4q\sqrt h/n^2 = O(q/F(n,\epsilon))$.

In all three regimes, Lemma~\ref{lem:metric-hybrid} yields $\TV(\Pi_{\mathcal A}(\mu_-),\Pi_{\mathcal A}(\mu_+))\le \Pr(F_{\mathcal A})+2h/(2n^{10}+1)<1/3$ whenever $q\le cF(n,\epsilon)$ for a sufficiently small absolute constant $c>0$ and large $n$, ruling out two-sided testers.
\end{proof}

\subsection{Contextual blow-up and adaptive metric testing lower bounds}

Unlike nonadaptive algorithms, an adaptive algorithm can use exact numerical distance answers to infer vertex coordinates and guide subsequent queries. To prevent global coordinate recovery, we blow up each vertex of a coarse robust face into $t$ clones and assign independent random coordinate bijections to each coarse tight-triangle context.

\begin{lemma}\label{lem:metric-blowup}
Let $(A_0,\mathcal H_0)$ be a robust face from Lemma~\ref{lem:metric-face} on $s$ vertices with slack and endpoint-margin bound $s^{-K}$ ($K\ge3$), and fix an integer $L>K$. Replace each coarse vertex by $t\ge2$ clones and pad with distance $3/2$ to $n\ge st$ vertices. For sufficiently large $n$, there is a random metric $A\in[1,C_0]^{\binom{[n]}2}$ with an edge-disjoint family $\mathcal H$ of $|\mathcal H|=|\mathcal H_0|\lceil t^2/2\rceil$ tight triangles (so $|\mathcal H|\ge\beta n^2/4$ when $st=n-o(n)$) such that all other triangle slacks and endpoint margins of selected edges are at least $n^{-(L+3)}$ and all entries are $O_C(\log n)$-bit rationals.
\end{lemma}

\begin{proof}
For each incidence of a coarse vertex with a coarse triangle in $\mathcal H_0$, independently choose a uniform bijection from its $t$ clones to $[t]$. Set $\tau=1/(100s^L)$ and $\alpha=\tau/t^2$. For each coarse triangle $(x,y,z)\in\mathcal H_0$ with long edge $xz$ and local clone coordinates $u,v,w\in[t]$, set $A_{XY}(u,v)=A_0(x,y)+\alpha(u-v)^2$, $A_{YZ}(v,w)=A_0(y,z)+\alpha(v-w)^2$, and $A_{XZ}(u,w)=A_0(x,z)+\frac\alpha2(u-w)^2$; keep all other cross-cell distances equal to $A_0$ and set within-cell and padding distances to $3/2$. Edge-disjointness of $\mathcal H_0$ ensures that each coarse pair belongs to at most one context.

On $(x,y,z)\in\mathcal H_0$, the slack of $(u,v,w)$ is $\frac\alpha2(u+w-2v)^2$, which vanishes precisely on the $\lceil t^2/2\rceil$ midpoint triples $u+w=2v$ and is at least $\alpha/2\ge 1/(200s^Lt^2)\ge n^{-(L+3)}$ otherwise. Because any two coordinates of a midpoint triple determine the third and distinct contexts are disjoint on coarse edges, the resulting fine tight triangles $\mathcal H$ are edge-disjoint. Every perturbation is in $[0,\tau]$ with $\tau\le s^{-K}/100$, preserving all unselected slacks and endpoint margins.
\end{proof}

\begin{lemma}\label{lem:metric-context-exposure}
Apply the signed noise perturbation of Lemma~\ref{lem:metric-noise} independently to the fine triangles of Lemma~\ref{lem:metric-blowup} and uniformly relabel $[n]$. For any adaptive $q$-query algorithm $\mathcal A$ with $2q\le n$ and any integer $1\le R<t$, the event $F_{\mathcal A}$ that $\mathcal A$ queries all three edges of some fine triangle in $\mathcal H$ satisfies $\Pr(F_{\mathcal A}) \le \frac{2qR^2}{t-R} + s\bigl(\frac{2eqt}{nR}\bigr)^R$ under either sign.
\end{lemma}

\begin{proof}
Generate the signed noise table in abstract coordinate space $[t]^3$ for each coarse context before sampling the bijections from clones to coordinates, and strengthen the oracle so that querying any edge in a coarse selected context also reveals the coarse cells of both endpoints and their local context coordinates (disclosing at most two new local coordinates per query).

First, conditioning on a fixed full template, the strengthened oracle answers are functions of the template vertex images, so Lemma~\ref{lem:lazy} embeds the $2q$ touched vertices into a uniformly random $2q$-set. By a union bound over the $s$ coarse cells, the probability that some cell has at least $R$ touched clones is at most $s\binom{2q}{R}\frac{(t)_R}{(n)_R}\le s\bigl(\frac{2eqt}{nR}\bigr)^R$.

Second, condition on the random seed, the abstract noise table, and the coarse cell partition of $[n]$. Because numerical edge weights in a context are deterministic functions of the disclosed local coordinates and the fixed abstract noise table, the conditional law of the remaining undisclosed coordinates in each incidence between a cell and a context remains uniform over the unused values in $[t]$. Stop the process if any cell reaches $R$ touched clones. Prior to stopping, each newly disclosed coordinate is uniform over at least $t-R$ unused values, while the other two cells of its coarse context have at most $R$ disclosed coordinates each and hence prescribe at most $R^2$ midpoint-completing values. A union bound over the at most $2q$ coordinate disclosures therefore bounds the probability of completing a midpoint triple before stopping by $2qR^2/(t-R)$.
\end{proof}

\begin{theorem}[Adaptive lower bound on a bounded range]\label{thm:metric-adaptive}
For every fixed $C>2$, there exist constants $c_C>0$ and $n_1(C)$ such that for all $n\ge n_1(C)$ and $0<\epsilon\le1/80$, any adaptive two-sided $\epsilon$-tester for general metrics on $[1,C]$ requires at least $c_C\sqrt{n/\log n}$ queries. The same lower bound holds for one-sided testers and unrestricted positive metric repairs, on inputs with $O_C(\log n)$-bit rational entries.
\end{theorem}

\begin{proof}
Set $\ell=\lceil\log n\rceil$, $s=\lfloor\sqrt{n/\ell^3}\rfloor$, $t=\lfloor n/s\rfloor$, and $R=20\ell$, and apply Lemma~\ref{lem:metric-blowup} with $K=3$ and $L=4$ followed by Lemma~\ref{lem:metric-noise} with $J=10$. For large $n$, we have $st=n-o(n)$ and $h=|\mathcal H|\ge\beta n^2/4$, so every instance in $\supp\mu_+$ is at distance $2h/n^2\ge\beta/2=1/40>\epsilon$ from $\mathcal M_n$. Moreover, $t-R\ge\frac14\sqrt n\,\ell^{3/2}$ and $2eqt/(nR)\le e/5$ whenever $q\le\sqrt{n/\ell}$, so Lemmas~\ref{lem:metric-context-exposure} and~\ref{lem:metric-hybrid} yield
$\TV(\Pi_{\mathcal A}(\mu_-),\Pi_{\mathcal A}(\mu_+)) \le 6400q\sqrt{\ell/n} + O(n^{-8}) < 1/3$
for $q\le c_C\sqrt{n/\log n}$ with $c_C>0$ sufficiently small.
\end{proof}

Finally, combining Theorems~\ref{thm:sharp-nonadaptive-metric} and~\ref{thm:metric-adaptive} for any fixed $C>2$ with Lemma~\ref{lem:clipping-equivalence} transfers both our nonadaptive two-sided lower bound across all proximity regimes and our adaptive two-sided lower bound directly to unrestricted metrics and pseudometrics.

\begin{theorem}[Lower bounds for metric testing]\label{thm:unrestricted-metric-lower}
On arbitrary nonnegative symmetric inputs, for all sufficiently large $n$ and $0<\epsilon\le1/80$, nonadaptive two-sided and adaptive two-sided $\epsilon$-testing of general metrics require, respectively,
$$
\Omega(F(n,\epsilon))
\qquad\text{and}\qquad
\Omega\!\left(\sqrt{\frac{n}{\log n}}\right)
$$
queries. Both lower bounds hold for either positive metric or nonnegative pseudometric repair targets, using hard inputs with $O(\log n)$-bit rational entries.
\end{theorem}

\begin{proof}
Fix $C=3$ in Theorems~\ref{thm:sharp-nonadaptive-metric} and~\ref{thm:metric-adaptive}. All instances in $\supp\mu_-$ are positive metrics in $[1,3]$, and Lemma~\ref{lem:clipping-equivalence} ensures that every instance in $\supp\mu_+$ has the same Hamming distance to $\mathcal M_n$ and $\overline{\mathcal M}_n$ as to $\mathcal M_n[1,3]$.
\end{proof}

\printbibliography

\end{document}